\documentclass[12pt]{article}
\usepackage{subeqnarray}
\usepackage[title]{appendix}
\usepackage{epsfig,amsmath,amssymb,mathtools}
\usepackage{enumitem,setspace}
\usepackage{mathrsfs}
\usepackage{empheq}
\usepackage{soul}
\usepackage{centernot}
\usepackage[usenames,dvipsnames]{color}
\usepackage[pagebackref=true, colorlinks=true]{hyperref}
\definecolor{redish}{rgb}{0.7,0.2,0.0}  
\definecolor{bluish}{rgb}{0.2,0.5,0.8}
\hypersetup{linkcolor=redish,          
                  citecolor=blue,        
                  filecolor=magenta,      
                  urlcolor=bluish}          

\newcommand{\nn}{\nonumber}
\def \({\left(}
\def \){\right)}
\def \[{\left[}
\def \]{\right]}
\begin{document}
\title{A logico-linguistic inquiry into the foundations of physics: Part I}
\author{Abhishek Majhi,\\  Indian Statistical Institute,\\Plot No. 203, B. T. Road, Baranagar,\\ Kolkata 700108, West Bengal, India.\\
	(abhishek.majhi@gmail.com)}%
\date{~}
\maketitle
\begin{abstract}
{\color{black} Physical dimensions like ``mass'', ``length'', ``charge'', represented by the symbols $[M], [L], [Q]$, are {\it not numbers}, but used as {\it numbers} to perform dimensional analysis in particular, and to write the equations of physics in general, by the physicist. The law of excluded middle falls short of explaining the contradictory meanings of the same symbols. The statements like ``$m\to 0$'', ``$r\to 0$'', ``$q\to 0$'', used by the physicist, are inconsistent on dimensional grounds because ``$ m$'', ``$r$'', ``$q$'' represent {\it quantities} with physical dimensions of  $[M], [L], [Q]$ respectively and ``$0$'' represents just a number -- devoid of physical dimension. Consequently, due to the involvement of the statement ``$\lim_{q\to 0}$, where $q$ is the test charge'' in the definition of electric field leads to either circular reasoning or a contradiction regarding the experimental verification of the smallest charge in the Millikan-Fletcher oil drop experiment.
	
	 Considering such issues as problematic, by choice, I make an inquiry regarding the basic language in terms of which physics is written, with an aim of exploring how truthfully the verbal statements can be converted to the corresponding physico-mathematical expressions, where ``physico-mathematical'' signifies the involvement of physical dimensions. Such investigation necessitates an explanation by demonstration of ``self inquiry'', ``middle way'', ``dependent origination'', ``emptiness/relational existence'', which are certain terms that signify the basic tenets of Buddhism. In light of such demonstration I explain my view of  ``definition''; the relations among quantity, physical dimension and number; meaninglessness of ``zero quantity'' and the associated logico-linguistic fallacy; difference between unit and unity. Considering the importance of the notion of electric field in physics, I present a critical analysis of the definitions of electric field due to Maxwell and Jackson, along with the physico-mathematical conversions of the verbal statements. The analysis of Jackson's definition points towards an expression of the electric field as an infinite series due to the associated ``limiting process'' of the test charge. However, it brings out the necessity of a postulate regarding the existence of charges, which nevertheless follows from the definition of quantity. Consequently, I explain the notion of {\it undecidable charges} that act as the middle way to resolve the contradiction regarding the Millikan-Fletcher oil drop experiment. In passing, I provide a logico-linguistic analysis, in physico-mathematical terms, of two verbal statements of Maxwell in relation to his definition of electric field, which suggests Maxwell's conception of dependent origination of distance and charge (i.e. $[L]\equiv[Q]$) and that of emptiness in the context of relative vacuum (in contrast to modern absolute vacuum). This work is an appeal for the dissociation of the categorical disciplines of logic and physics and on the large, a fruitful merger of Eastern philosophy and Western science. Nevertheless, it remains open to how the reader relates to this work, which is the essence of emptiness.}

\end{abstract}
\newpage
\tableofcontents
\newpage
\section{Introduction}
  When it comes to writing down theories for written explanations of observed phenomena, then the inherent consistency of the language, in terms of which the theory is being written, is necessary. And, as far as the consistency of  reasoning based on mathematics is concerned, arithmetic founds the root of all \cite{poincareq1,poincare,fregeq1,frege1,weyl,weyl1,ifrah1,ifrah2}. That is to say, the essence of numbers lies at the heart of the mathematical part of reasoning because  the basic operations in mathematics like addition, multiplication, etc., only numbers take part \cite{lilavati,peano}\footnote{In ref. \cite{peano}, Peano considered the words ``real positive number'' and ``quantity'' synonymously. This synonymy is questioned in this work. Further, in usual everyday life for practical purpose and for physicists, the usual arithmetic  discussed in ref.\cite{lilavati} suffices and the formal representation in terms of logical symbols are not necessarily required, except by professional logicians, computer scientists, etc.}. However, for the physical interpretations of the observations,   verbal expressions\footnote{Since I am writing to be read by the reader, rather than speaking to be listened by the listener, ``verbal expressions'' actually appear as ``written expressions'' in terms of the concerned metalanguage, which is English here.} become indispensable. Therefore, the inherent consistency of the language depends on how well the mathematical and the verbal expressions complement each other.
  
  \subsection{Revisiting the primary lessons of physics{: physical dimensions and units}}
   The primary lesson in physics, which makes the subject distinct from mathematics, is that of {\it physical dimensions}\footnote{I have used the adjective ``physical'' so as to distinguish from ``degrees of freedom'' which are also called ``dimension'' e.g. three dimensional space, two dimensional space, etc. However, this brings in the negation of ``physical dimension'', which may be called ``unphysical dimension''. I plan to address such issues on a separate occasion; {\color{black}however, such logico-linguistic obstacles plague the language of the logico-linguist himself e.g. see ref.\cite{companion}.}}, represented by symbols like $[M],[L],[T]$ to express the thought of physicality associated with the verbal expressions like ``mass'', ``length'', ``time'' respectively. Although such clarifications were not provided by Newton \cite{principia1,principia2}, the use of such symbols got a formal recognition through dimensional analysis over the years \cite{maxwell,bridgman, dimhis}\footnote{I may note that Maxwell used the symbols like $[M],[L],[T]$, etc. to denote units in ref.\cite{maxwell}.}. Such physical dimensions provide the ground for the physicist to make the distinction between number and {\it quantity}. For example, the expressions ``mass of an object'' and ``volume of an object'' are used to express the thought of the respective quantitative aspects of the concerned object. The expressions differ with respect to  physical dimension. Replacing the words ``mass'' and ``volume'' by ``number'' erases this difference and does not serve the purpose meant to be served by the respective words ``mass'' and ``volume''.  
   
   The other primary lesson of physics is that of {\it units}, which are stated through verbal expressions like ``kilogram'', ``metre'', ``second'', etc. {(\color{black}conventionally called {\it unit names}\cite{nist})}. Such verbal expressions are usually further abbreviated as symbols for typographical convenience{\color{black}(conventionally called {\it unit symbols}\cite{nist})}.  
    These units are {\it not} numbers by themselves. Rather, they symbolize conventional or chosen standard  quantities with which other quantities of the same physical dimension can be measured in terms of positive whole numbers\footnote{For example, Einstein expressed such view while explaining the foundations of relativity, especially on page no. 4 of ref. \cite{einphil2}. Also, such view was shared by Russell \cite{russellnum} (also see Appendix \ref{appb}).}. This provides the essence of measurement in physics.

   Such primary lessons are somehow forgotten while writing down the basic equations in physics. Rather, physics is written in a reckless fashion, where the concepts of {\it quantity} and {\it unit} are identified with {\it number} and {\it unity} respectively, without any reasonable explanation \footnote{Ref. \cite{essay} is an exception.}.   For example, ``force ($F$) = mass ($m$) $\times$ acceleration ($a$)'', ``momentum $(p)$ = mass $(m)$ $\times$ velocity $(v)$'', ``force ($F$) $\propto$ product of two masses ($m_1\times m_2$)'' etc. are expressions which showcase arithmetic operations with  non-mathematical symbols\footnote{Mathematics deals with the operations like addition, multiplication, etc. with numbers represented as numerals and algebraic symbols. So, it is the arithmetic nature that forms its heart. Only the use of those mathematical results, for  interpreting practical observations, necessitates the use of words and symbols to represent physical experience of  human being e.g. length, mass, volume, etc. } i.e. $F, p, m$, etc. are not just numbers. Here, one can provide a reasonable argument by saying that, ``In the expression $F=ma$, the conveyed meaning is the following: $m=n_m m_0, a=n_a a_0, F=n_F F_0$, where the $n$-s are numbers\footnote{It is extremely tempting to write ``pure numbers'' in place of ``numbers'' so as to emphasize the absence of any association of physical dimension to any number. However, then a question may arise whether there is a notion of ``impure number''. I avoid such meaningless traps of words by not using unnecessary adjectives.}, $m_0, a_0, F_0$ are units  and the equation actually implies $n_F = n_m n_a$.'' However, this argument is not satisfactory enough because a postulate like ``$F_0=m_0\times a_0$'' is  necessary
   to write down the equality of $n$-s. Moreover, the sudden appearance of $F_0,m_0,a_0$ in the scenario requires further explanations. So, the argument that seemed apparently reasonable, is after all fraught with the same problem of arithmetic operations with non-mathematical symbols. Thus, physics is foundationally non-arithmetic in nature \footnote{Poincare did express his concerns about the non-arithmetic foundations of physics, particularly on page no. 6 of ref.\cite{poincare}. Also,  Dedekind would go as far as writing, on page no. 1 of ref. \cite{dedekindnum}, that the geometrically founded  ``introduction into the differential calculus can make no claim to being scientific'' and would decide to keep searching for an ``arithmetic'' foundation of the same.}. 
   
   {\color{black}Now, it is true that in a standard source book of the present day which discusses units, namely ref.\cite{nist}, it has been clearly stated on page number 29 that,\vspace{0.1cm}

   {\small``{\it Unit symbols are mathematical entities.... In forming products and quotients of unit symbols the normal rules of algebraic multiplication or division apply.}''}\vspace{0.1cm}
   
   However, such declarations do not come along with the justification regarding why units, which are not numbers (unit symbols are not numerals), can be used as numbers so that arithmetic operations can be carried out with the corresponding symbols. Further, it has also not been clarified why addition and subtraction are not allowed if multiplication and division can be done. In a nutshell, the doubt remains regarding the use of symbols, which do not represent numbers (physical dimensions and units), but are used as numbers to write the theories of physics.
   }

Strangely enough, even today, I doubt whether such basic issues regarding the foundations of physics are even recognized in general and in particular, while considering questions regarding the foundations\cite{massgap,navsto}. {\color{black} Even stranger is the fact, as it appears to me, that the philosophically driven scientists, who try to look beyond the convention regarding quantity, unit, measurement, etc., do not address this simple query for which no sophisticated jargon or notations or technicalities are required, in spite of the development of the respective literature in its own right e.g. see ref.\cite{michell,stanmea,wolff} and the references therein.}

   \subsection{Mathematics and mathematical logic: no physical dimension}
    Foundations of mathematics have been dealt with great endeavours \cite{heijen,bragg,gilliesmath}, leading to axiomatic set theory \cite{fraenkel,cantor1,cantor2}, albeit fraught with inherent inconsistencies (like Russell's paradox, etc.\cite{pm1}) and also leading to the identifications of the incompleteness of formal arithmetic systems\footnote{ There are certain propositions in an arithmetic system which are neither provable nor disprovable within the system. That is, there can be a proposition $P$ such that neither $P$ nor its negation $\neg P$ is provable within the system. Such a proposition $P$ is called undecidable proposition. If Aristotle's sense of either truth or falsity of $P$ is considered, then there can be $P$ such that ``both $P$ and $\neg P$'' can be true and also ``neither $P$ nor $\neg P$'' can be true, which implies the limited validity of the law of excluded middle.} \cite{goedelincom}. While such developments were based on  Aristotle's law of the excluded middle \cite{aristotlepa}, further developments of the foundations of mathematics beyond such premises have taken place leading to multi-valued and fuzzy mathematical logic\cite{logichandbook8,multifuzzy}. Although such efforts have dealt with mathematical reasoning, involving numbers and arithmetic operations, and also with the algebraic structure of the logico-symbolic expressions of thought, however, such literature do not address the questions regarding the non-mathematical symbols which represent physical dimensions in physics. As a result, throughout the literature concerning physics and mathematics, till date, no distinction has been made between quantity and number, and hence, unit and unity\footnote{I find an exception  in the attitude of Frege \cite{fregeoneunity}, given my limited knowledge of the scientific literature irrespective of any category.}. While such development of logic has led to the modern sophisticated technologies, such as intelligent systems\cite{fuzzy1,fuzzy2}, etc., which were only dreamt of during the days of Leibniz \cite{logichandbook3}, it is hardly deniable that physics, as began to get formalized by Newton \cite{principia1,principia2}, holds the key for the usefulness of mathematical logic for practical applications. This is because machines need to be built with the knowledge of physics to implement the knowledge of mathematical logic.

    However, the logician has the right to be curious about the logical soundness of the basic structure of reasoning, the definitions used and the language in terms of which thoughts are expressed by the physicist. Most certainly the logician can/should demand a clarification about the logical status of physical dimensions due to the following reason.  Although the concept of physical dimension is a priori distinct from the concept of number, the same is used as number while performing arithmetic operations with quantities, which formally began with Newton  \cite{principia1,principia2}, and also while performing dimensional analysis (e.g. see ref.\cite{bridgman,dimhis}), without which theories of physics become meaningless writing. Considering the law of the excluded middle, the logician should not be able to understand how any physical dimension is used both as number and not number to write theoretical explanations in physics. 
  
    Consequently, it should not be hard to find apparent inconsistencies among expressions of thoughts made in terms of verbal language and the corresponding {\it physico-mathematical} expressions in terms of which the equations of physics are written (I use the word ``physico-mathematical'' so as to justify the involvement of physical dimensions alongside  numbers).
     Unfortunately, such basic questions regarding the foundations of physics have never been raised by the logician in course of development of science \cite{heijen,logichandbook3,nagel,popper,wittgenstein}(also see ref.\cite{companion}) or by the physicist who intends to focus on the logical structure of science \cite{duhem,logicphysics}.

   \subsection{Aim and structure of this work}
   The aim of this work is to revisit the basic premises with which physicists work so as to investigate how reasonably the verbal expressions of thought are converted to physico-mathematical ones {\color{black}and how truthfully the experiences are conveyed through such expressions. Nevertheless, the motivation behind such a work has germinated through several layers of inquiries regarding my own understanding of some very basic concepts of physics which, however, may be tagged as ``an illiterate's thought'' by the literate reader unless he himself goes through such inquiries with a scholarly mindset unshackled from the grip of any prevalent dogma. For example, the prevalent habit of the physicist is to write statements like ``$r\to 0$'', ``$m\to 0$'', ``$q\to 0$'' where $r, m, q$ have the physical dimension of length, mass, charge, respectively and $0$ is a number on the right hand side of the respective statements. I find such statements to be unacceptable because of dimensional mismatch. Moreover, I consider it as an untruthful dogma because the physicist himself considers dimensional match (homogeneity) as the most basic criterion for consistency of any physico-mathematical expression is physics. If the reader does not find any problem in such statements as ``$q\to 0$'',etc. and accept their validity, then this work must appear to him as a worthless piece of writing. However, the reader, who dares to encounter such a widely prevalent dogma, may find this work worth reading.

   	My inquiry germinated from the physicist's definition of electric field, that involves a statement ``$q\to 0$'', and the problems associated with such a definition from the logician's perspective of the word ``definition'', if I try to verbally interpret the significance of the statement ``$q\to 0$'' in accord with what I find in standard texts.  Consequently the issue demanded a necessity of a clarification of the difference between the concepts of ``quantity'' and ``number'', ``unit'' and ``unity'', which in turn necessitated a clarification of the logical status of physical dimension. This is the reason for what I have written in the previous two sections. However, a clarification of the use of physical dimension to write down theories of physics required me to go beyond the foundations of logic. What I encountered is that, the attitude of making self-inquiry which leads to the realization of the middle way and some associated subtle aspects of reasoning, which are the basic tenets of Buddhism, to be missing in both the physicist's and the logician's approach to their respective subjects. 
   	
   	Therefore, as far as the structure of this work is concerned, I begin, in section (\ref{nonbody}), with some clarifications of the terms such as ``self-inquiry'', ``middle-way'', ``emptiness'', ``dependent origination'' in my own words along with a simple demonstration through drawing. Since the literature of physics (and logic), as far as my knowledge is concerned, is devoid of such discussions, I present my work in simple language without using any sophisticated notations of mathematical logic,  except a few elementary symbols, so that the work is readable to any person irrespective of being a physicist or a logician or a philosopher or a mathematician.   
   
    In section (\ref{definition}), I clarify my stance regarding the term ``definition'', in light of self-inquiry, along with simple demonstrations. I also explain in short how my view neither matches completely with that of the logician who seeks completeness of truths through definitions nor matches completely with that of the physicist who relies on the operational value of definitions with little attachment to logic. Nevertheless, the spirit in which I consider the term ``definition'' can be better understood through further demonstrations and critical analyses of known definition of electric field in terms of physico-mathematical expressions which I provide in later sections.
    
    In section (\ref{qpdn}), I provide a thorough discussion regarding quantity, physical dimension and number. I begin with the definition of quantity in terms of physical dimension and number along with an explanation of dependent origination of the three concepts through self-inquiry. I also clarify how  ``zero quantity'' leads to a logico-linguistic fallacy, which nevertheless plagues the literature of physics, and the essence of emptiness in regard of the existence of quantity. Then I explain how the logical status of physical dimension can be explained through the middle way by providing an example of addition of two quantities.
   
   In section (\ref{unitunity}), I explain the difference between unit and unity in light of the definition of quantity. Further I discuss, in this context, how the views of the theoretical physicist and the experimental physicist differ and why there is a profound mismatch between the experience of the physicist and the corresponding written expressions. The discussion concerns the truthfulness or morality of the physicist's expressions. I provide such clarifications in order to demonstrate how the physicist lacks the attitude of self-inquiry.
   
   In section (\ref{physdeffield}), I discuss the definition of electric field as given by the physicist in available texts and associated objections from the logician's perspective. The objections arise due to the involvement of ``zero quantity'' as a limiting condition and the associated verbal interpretations. Considering the available explanations in the literature, I explain how the verbal interpretations either depends on circular reasoning or remains incomplete due to an unresolved contradiction. The contradiction arises due to the experimental verification of the smallest charge which gets into conflict with the requirement of arbitrarily small charge for the definition of electric field. This concerns the oil drop experiment performed by Millikan and Fletcher\cite{millikan,fletcher}. 
   
   In section (\ref{pmvs}), I provide a physico-mathematical analysis of the verbal statements in terms of which the electric field has been defined in two classic textbooks, namely, those of Maxwell\cite{maxwell} and Jackson\cite{jackson}. The analysis reveals how a truthful conversion of  the verbal statements to the corresponding physico-mathematical ones could have led to a differently structured expression for the electric field than what is known today. Nevertheless, such analysis reveals the necessity of a postulate that can be considered as a premise to define the electric field. So, I declare the postulate that is necessary for the definition of the electric field, which is a straightforward consequence of the definition of quantity of charge that I have discussed in section(\ref{qpdn}). I clarify further how such a postulate resolves the contradiction that concerns the oil drop experiment. In the process, I introduce the concept of undecidable charges i.e. the charges which are neither verifiable nor falsifiable either in theory or by experiment. Rather, these charges act as the middle way -- the founding premise -- on which rests the logical consistency of the theory and the observation concerning the oil drop experiment. Further, I specifically analyze two verbal statements made by Maxwell prior to the definition of electric field in ref.\cite{maxwell} to explain how he could have written his theories had he made conversions to physico-mathematical expressions in light of self-inquiry and dependent origination. I explain how Maxwell's statements suggest that charge and length can possibly considered to be of same physical dimension and the concept of vacuum is only relative, in contrast to the presently accepted concept of ``absolute vacuum'', which is however the essence of emptiness or relational existence.
   
   In section (\ref{summary}), I conclude with a summarizing discussion of the my work with an intent to showcase why this work can be considered as an appeal for a fruitful merger of Western science and Eastern philosophy. 
}

{\section{Neither the logician nor the physicist -- being `nobody'}\label{nonbody}

	When the physicist claims to be logical, it is worth noting that according to Boole, while using symbolic expressions there needs to be,  ``{\it in the same process of reasoning}'',  maintained a ``{\it \underline{fixed} sense or meaning of the symbols employed.}''(see page no. 6 of ref.\cite{boole}). Simply based on this single most criterion the logician can argue that physics is not logically well founded because the  written expressions of physical dimensions, through the corresponding symbols, apparently carry different  meanings. However, the physicist can defend his stance on the ground that  physics has led to the development of technologies which after all justify the purpose of doing science i.e. to have practical applications. In that case, the logician's argument does not appear  to have any value if practical application is considered as a necessary, sufficient and ultimate criterion for science. On the other hand, the physicist's argument seems untenable because the modern technologies do get the richness through the use of computers which run on logical programming \cite{turing}. Of course then the question arises whether the categorization, as ``logician'' and ``physicist'', itself has any ultimate value; or, it is a middle path, that tends to adopt a non-categorized approach towards scientific queries, is  of utmost use to mankind. A reasonable justification for such a thought is the evidence that both the literature of logic and physics have developed in their own rights, which only shows that what becomes useful to mankind, irrespective of any categorization, serves the purpose of doing science. With such a mindset, I must walk along such a middle way by being neither the logician nor the physicist -- a nobody-- to present this discussion.

	To express my own attitude with which I proceed, I must explicate a few words which appear in the context of Buddhism viz. self-inquiry,  middle-way, emptiness, relational existence or dependent origination. Since I have not been able to find, in the literature, such inquiry regarding physics or science in general (which I intend to present here), I prefer to explain the essence of such words in my own way rather than referring to any particular texts\footnote{{\color{black}Buddhism has metamorphosed through several stages after the death of the Buddha and different interpretations of Buddhism can be found in plethora of modern texts e.g. see refs.\cite{siderits,garfield,nayak,stre,bodhi} and many more. Such texts are devoid of any investigation regarding the foundations of physics. However, there are instances where modern authors draw connection between modern physics and Buddhism e.g. see {\it Part 3: Buddhism and the Physical Sciences} in ref.\cite{wallace}. Such texts are devoid of any logico-linguistic inquiry, regarding the foundations of physics, that I intend to present here. So, I must refrain from making unnecessary comments regarding such texts.}}. What I write in the following subsection is a manifestation of my own realization that I find very similar to the basic tenets of Buddhism \footnote{{\color{black}Certainly it will be quite an act of deliberate ignorance on my behalf if I do not mention Indian philosophy as a whole in the context of such a discussion and restrict my citations to Buddhist texts only. This is because the discussions regarding the words ``self-inquiry (atmavichar)'', ``emptiness (shunyata)'', ``middle way (madhyamarg)'', ``dependent origination (pratityasamutpada)'' concern very subtle demonstrations concerning the Vedas, the Upanishads, etc.(e.g. see ref.\cite{surendranath} and other volumes, ref.\cite{radhakrishnan} and other volumes, ref.\cite{sinha}, ref.\cite{potter2} and other volumes and many more that I do not cite here.). However, I have found my viewpoints closer to those which I find in Buddhism, especially the denial of an eternal self e.g. see ref.\cite{bodhi}.  I have ignored such discussion in the present context as it will effect in too much digression as far as the present work is concerned.}}. 
}

{\subsection{Self-inquiry, middle way, emptiness, dependent origination: a demonstration with drawing}\label{simwdo}
	I draw a closed line with my pencil on paper -- see fig.\ref{linemiddleway}. The question I ask myself whether I have a complete realization of what is area (i.e. the region on the paper enclosed within the closed line) and what is not area (i.e. the region on the paper that is not enclosed within the closed line). There are two choices that I can make to answer the question.
	\begin{enumerate}
		\item I assume that I have an understanding of the whole sentence ``I draw a closed line with my pencil on paper'' and proceed with this assumption.
		\item I investigate each of the words in the sentence ``I draw a closed line with my pencil on paper'' and seek an explication of each and every term i.e. what are the meanings of ``pencil'', ``paper'', ``line'', ``draw'', ``closed'' and most importantly who/what am``I''.
	\end{enumerate}  
	\subsubsection{The ``I'' (silence): neither the self nor the non-self, neither the body nor the mind, nor the consciousness}\label{I}
	The second choice is directed towards going beyond language and having direct experience to understand how truthful I am while expressing my experience. This is an example of self-inquiry i.e. I investigate my own truthfulness.  So, I start with my realization\footnote{{\color{black}``My realization'' or ``my consciousness'' or ``my awareness'' -- I use all these three alternatives in an equivalent sense.}} of the self and the non-self, where the self is that which I attempt to refer to as the ``I''. However, I can not realize the ``I'' through any mode of expression like moving my body, speaking some language or through any act in general. Rather, I can only end up referring to what is mine i.e. my body and other material possessions like my pencil, my paper, etc. Being a bit more subtle I can refer to my mind as I think. But I can not think of a complete distinction between the mind and the body as I struggle with the doubt whether I can have the realization (or become conscious or aware) of  my mind without my body and the realization of my body without my mind. Even after such doubt, I think I am aware of the self through both the mind and the body. However, if there is doubt in the mode of realization or of the arousal of consciousness (mind-body\footnote{\color{black}May be Bodhi would write ``mentality-materiality'' instead of ``mind-body'' e.g. see ref.\cite{bodhi}.}), then I wonder how I can be certain about the realized or conceived distinction between the self and the non-self. Therefore, my realization of or consciousness of my own existence is neither that of self nor that of non-self but a middle way between the two. The reference to the self and the non-self occurs neither through only the body nor through only the mind, but I require both to think and express. Thus I realize, through self-inquiry, the essence of the middle way i.e. the realization or the consciousness or the  awareness, that gives rise to the categories ``body'' and ``mind'' and the categories (mind-body) that provides the mode of the realization or the consciousness or the awareness. The ``I'' remains out of grasp and hence, inexpressible. Consequently, I conclude that all expressible truths are incomplete  -- there is always an element of doubt retained through the inexpressible ``I''. The (previous) statement applied to itself becomes self-destructive, but through a never ending series of steps of application to itself because each step contains an element of doubt which renders complete self-destruction impossible. I must make a choice to cut somewhere and ignore a part of the whole truth to make the rest a conventional truth with which I can work. But, once I cut, it becomes incomplete truth with which, however, I can work. Since such choice is dependent on the human being who reasons, the conventional truth becomes relative. In what I write, I choose to ignore the truth related to the ``I'' and write about what I can express through any writing. While making this choice of ignorance, I accept the most basic contradiction that I express the inexpressible by writing the ``I''. In case the reader makes this choice without making self-inquiry, which is also a choice, the bunch of sentences that I have written in this section, starting from the second line until the previous one, may appear to the reader as ``non-sense'' or ``rubbish'' or ``silly'' writing, or the reader may be a bit more respectful and judge it as ``mystic'' or ``pseudo-scientific'' writing. This is a judgment that has been made based on a relative truth -- a relation that the reader holds with what I have written and therefore, it depends partly on the reader's perception. However, any categorization of, such a conclusion drawn by the reader, by a second reader as ``false judgment'' is also a relative truth because this second reader is possibly aware of self-inquiry. The debate thus goes on with never ending doubt. To stop the debate, there should be silence.

	But, to do science I need to express and my expressions are manifested through some mode of expression, the mode being accessible to me i.e. mine. Therefore, the basic act of ignorance that I perform in order to do science is to consider the ``I'' as the indescribable premise of science. 
	Importantly, such ignorance or exclusion of the premise is a practice that involves my modes of perception and expression as well.  I have to exclude my eyes when I say that ``I can see'', because I can not see my eyes with my own eyes -- at least I need a mirror. So, I should have said that ``I can see, except my eyes which let me see''. The truth of my vision is only valid up to my eyes and does not include the truth of my eyes. Further, I have to exclude my fingertip when I say that ``I can touch'', because I can not touch my fingertip with that particular fingertip -- at least I need a different body part. So, I should have said that ``I can touch with my fingertip, except the particular fingertip itself.'' In general, I should say, ``I can observe, except the mode of observation that let me observe.'' Then, in the process I have to exclude my mode of observation, which is neither my body nor my mind, nor my consciousness that makes me aware of both my mind and my body. Rather, it is a subtle combination of all,  mind-body-consciousness, that constitutes the mode of observation. Then, who/what is the observer? It/he\footnote{{\color{black}Although I write ``he'', the ``I'' is beyond any such category. I am restricted by the limitation of English language.}} is the one who/which uses the mode of observation to observe i.e. the ``I''. Observation of the observer leads to silence which essentially is equivalent to saying that the ``I'' is inexpressible. So, the observer decides what is observed and what is not observed through the mode of observation which act as the middle way.  Therefore, the whole truth of observation, that includes perception and expression, contains the mode of observation as the middle.
	
	
	
	\subsubsection{The closed line: neither the area nor the not area}\label{closedline}
	Now, I come to the first choice i.e. I assume that I have an understanding of the whole sentence ``I draw a closed line with my pencil on paper'' and proceed with the assumption. This ``assumption'' is nothing but the choice of ignorance of the doubts cast by the second choice which I have discussed until the last paragraph. Thus the whole truth of the scenario includes both the two choices and it is the ignorance of the second that lets me work with the first choice. In this act, I express my thought of area and not area by drawing with a pencil. The truth of the area  and not area relies on the closed line. But the line itself is neither the area nor the not area and rather leads to the realization of both the area and the not area. The truth of the drawing includes the whole i.e. the area, the not area, the closed line which itself is neither the area nor the not area but it leads me to the realization of both the area and the not area -- see fig.\ref{linemiddleway}.
	\begin{figure}[hbt]
		\begin{center}
			\includegraphics[scale=0.40]{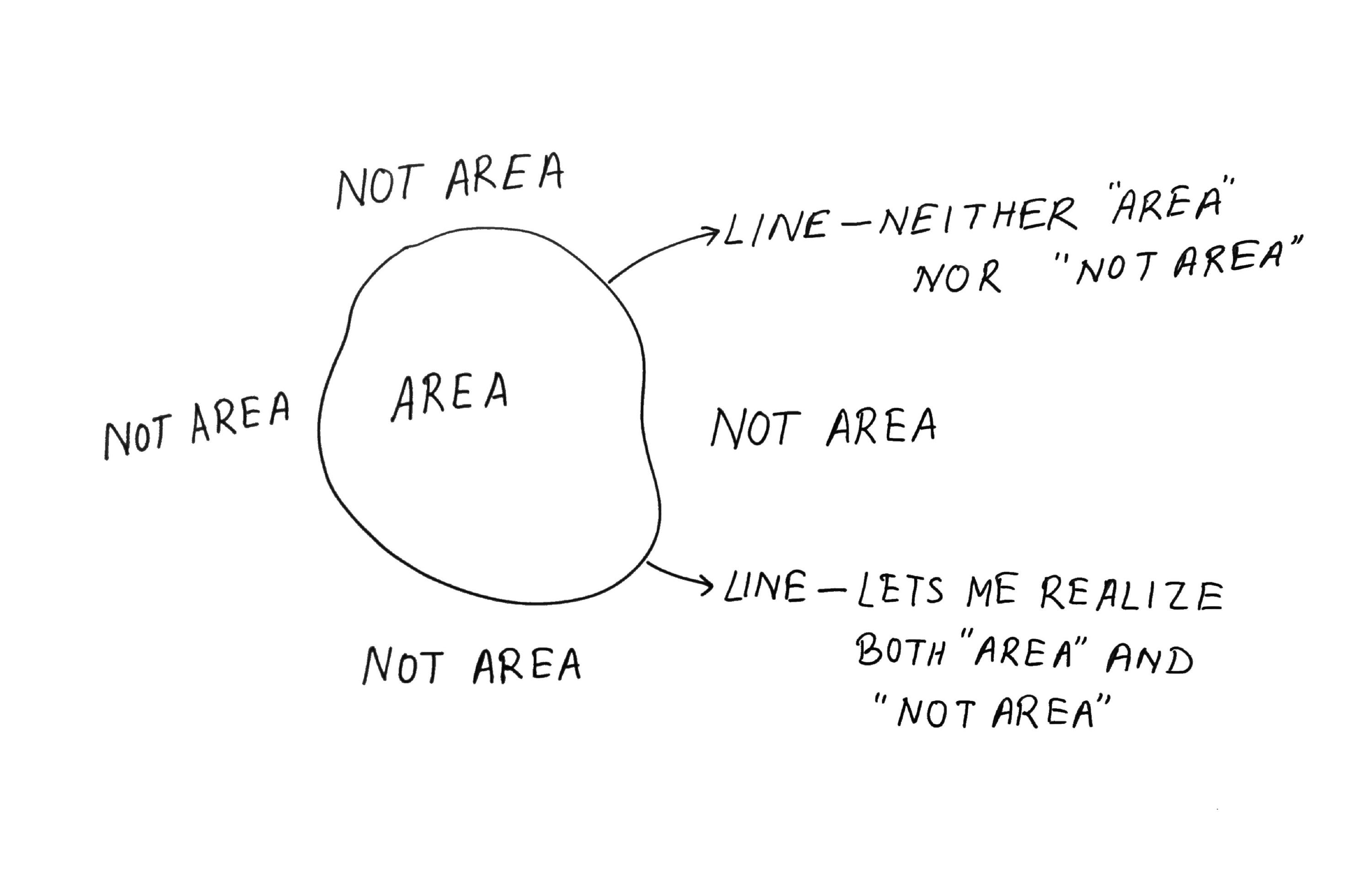}
			\caption{\label{linemiddleway}The line itself is neither ``area'' nor ``not area''. But I can realize both ``area'' and ``not area'' due to the visual perception of the line. }
		\end{center}
	\end{figure}
	The line can be made thinner by choosing a sharper pencil i.e. the middle way can be refined to make the distinction between ``true'' and ``false'', or two categories in general, sharper. But, a complete removal of the line renders the concepts of ``area'', and hence ``not area'', meaningless -- see fig.\ref{nolinemiddleway} i.e. the middle way can not be completely removed in practice or in the act of addressing the whole truth.
	\begin{figure}[hbt]
		\begin{center}
			\includegraphics[scale=0.40]{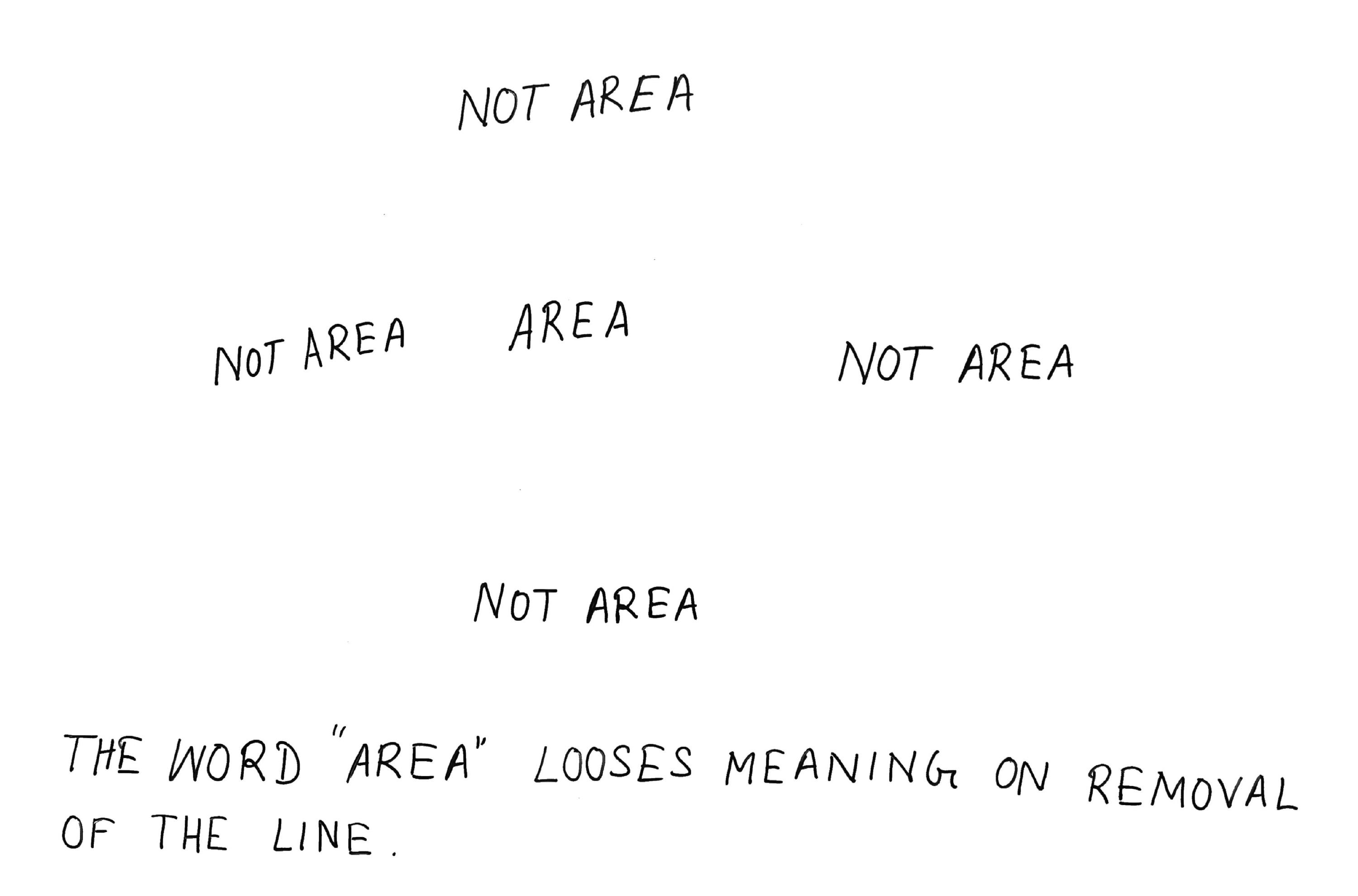}
			\caption{\label{nolinemiddleway}If there is no line I can not make the judgment of what is area and what is not area. The word ``area'' becomes meaningless.}
		\end{center}
	\end{figure}
	
	However, if I exclude the closed line by assumption, what is left with me is a complete distinction -- either the area or the not area. 
	This is a choice of ignorance or exclusion which lets me work with a certain assumed truth without referring to the context which lets the truth be. Here, the closed line is the middle way, in the limits of which fall the area and the not area. The complete distinction between area and not area, carries the sense of (classical) logic \footnote{{\color{black}Such self-inquiry is not done by the logician who tries to claim that logical truths must be complete e.g. see ref.\cite{companion} to compare with Frege's metaphor. Also, such self-inquiry is not done by the physicist who dreams of a final theory\cite{weinberg} or a theory of everything \cite{hawking} or provides a complete guide to an understanding of reality\cite{roadtoreality}. Consult  ref.\cite{companion} for further comments.} }. In other words, it is the exclusion of the middle way, by choice, that lets me work with logical truths and hence to make judgments. This is what I know as the `law' of excluded middle (way) and due to the above demonstration I may consider it as the `choice' of excluded middle (way).} 

\section{Meaning of ``definition''}\label{definition}
{\color{black}There is always an element of doubt, in any expression, that can be understood through self-inquiry. Such is the case with the title of this section. It is only meant for convenience because it is just an attempt to convey the phrase `definition of ``definition''~'. The logician would consider the nomenclature as circular reasoning and hence, fallacious. However, the circle can be broken by casting doubt in the completeness of such a conclusion. To do so, an explanation of the term ``definition'' must itself contain an element of doubt, by declaration. This is what I intend to explain.}

The physicist expresses his thoughts in terms of words like ``mass'', ``length'', ``time'', etc.  The meaning of such terms can not be explained in terms of more elementary ones, {\color{black} by assumption}. Rather, such terms are considered as the basis of further terms. On necessary grounds for the upcoming discussion, I provide the following clarification for the word ``definition'':\vspace{0.1cm}

  {\color{blue} An expression of a complex thought, made in terms of elementary premises which are expressions of elementary thoughts, is what I call {\it definition}. Any definition is written in terms of physico-mathematical expressions along with verbal expressions {necessary for resolving doubts.}}\vspace{0.1cm}

 Since this discussion concerns physics, the above meaning of definition is contextual and should be sufficient for the present purpose. {\color{black} Also, the clarification of the word ``definition'' is better understood through the demonstrations that I provide throughout this discussion because a complete clarification of the word ``definition'' is not possible. It always retains an element of doubt because the words ``meaning'', ``clarification'', ``definition'' are all synonymous in my understanding. So, my motive is to proceed through reasonable arguments to define certain terms for the present discussion which are useful for resolving some further doubts that must arise in due course. Nevertheless, the reader who seeks a complete clarification of the word ``definition'' is suggested to consult Appendix (\ref{logicians}) for some comments and particularly ref.\cite{companion} to see how the seekers of completeness are unaware of self-inquiry and as a result they have remained unaware of elementary issues regarding the foundations of physics.}

  I explain how this meaning  can be observed. I consider the following {\it elementary} physical dimensions viz. mass dimension $\equiv [M]$,  length dimension $\equiv [L]$, time  $\equiv [T]$, charge dimension $\equiv [Q]$, as the premises to define the others, which I call {\it derived} physical dimensions. For example, if I represent velocity dimension by the symbol $[V]$ then it is defined as $[L][T]^{-1}$; if I represent acceleration dimension by the symbol $[A]$, then it is defined as $[L][T]^{-2}$; if I represent force dimension by the symbol $[F]$, then it is defined as $[M][L][T]^{-2}$ and so on. In a nutshell:
\begin{eqnarray}
[V]:=[L][T]^{-1},\quad [A]:=[L][T]^{-2},\quad [F]:=[M][L][T]^{-2}.\label{derdim}
\end{eqnarray}
These examples do not contain any verbal expressions because these are  simple ones, {\color{black}but only if I do not make further self-inquiry. This is because, in order to be more precise about mapping of words to physico-mathematical symbols, I may write $[Q]:=[D]_{charge}$ where ``$[D]$'' is the physico-mathematical symbol for the words ``physical dimension'' and the lower index declares the ``type''. Similarly, $[M]:=[D]_{mass}, [L]:=[D]_{length}$ and so on. In this way the role of verbal expressions becomes apparent. Nevertheless, I shall not use such symbols and proceed with the usual accepted symbols like $[M], [Q], [L], [T]$, etc. by stopping self-inquiry beyond this point. }

Such definitions, as (\ref{derdim}), are the working principles in dimensional analysis \cite{maxwell,bridgman,dimhis} and such clarifications may appear naive and possibly useless nitpicking to the physicist, especially to the modern and sophisticated one, until the question is raised by the logician: {\it Are physical dimensions ``numbers''? If yes, then why are there so many symbols for representing one concept ``number''? If not, then what is the meaning of writing the arithmetic operations with such symbols to define the derived dimensions?}. 

Before going into discussions regarding the above queries let me point out that my view of definition does not match completely with either that of the logician or of the physicist, which I explain in the following sections.

\subsection{Comparison with the logician's view}\label{logicdef}
{\color{black} When the word ``definition'' becomes the matter of discussion itself, then it becomes generally the concern for  the logician (the philosopher, the logico-linguist). My view of ``definition'' both matches and differs with the general perspective of the logician. I explain the similarities and the differences in short, in what follows, to provide an overview. A detailed comparison is redundant for the present purpose. The interested reader can consult Appendix \ref{logicians}, and ref.\cite{companion} for further clarifications, to understand the redundancy.}

I find my view regarding the meaning of definition very similar to that of Tarski\footnote{\color{black} I have mentioned only Tarski's work regarding the meaning of ``definition'' with an aim to capture the logician's attitude in general. However, there are several other logicians who have had their views in this particular issue. Certainly such comparisons are also desirable if I want to make a statement regarding the logician in general. Nevertheless, given the fact that there has been no contribution of such logicians in the field of physics, such a discussion in the main section shall only add to unnecessary digressions. Also, I avoid the use of the specific jargon used by the logician in mathematical logic that involves sophisticated notations which often glosses over the essential shortcoming of reasoning that can be more directly demonstrated. For further comments regarding such issues the reader may consult Appendix \ref{logicians} and also ref.\cite{companion}.}(see page no. 152 of ref.\cite{tarskiundef}):\vspace{0.1cm}

{\small ``{\it The question how a certain concept is to be defined is correctly formulated only if a list is given of the terms by means of which the required definition is to be constructed. If the definition is to fulfill its proper task, the sense of the terms in this list must admit of no doubt.}''}\vspace{0.1cm}

 Here, physical dimensions like $[M],[L],$ etc. can be considered as the ``terms'' which fill Tarski's ``list'' from which the derived physical dimensions are ``constructed''.  In the above basic examples of constructing derived physical dimensions, the ``definition'' does ``fulfill the proper task'' because the ``terms in this list'', namely $[M],[L],$ etc.,  ``admit of no doubt'' by postulate or assumption -- those are the basic elements of thought expressed in terms of non-mathematical symbols. I call physical dimensions like $[M],[L],[T],[Q],$ etc. as {\it elementary physical dimensions} from which other physical dimensions can be derived.  

The question definitely arises whether there can be a complete list of such elementary premises from which all definitions in physics can be constructed {\color{black} and my view differs from that of Tarski, and the logician in general, due to the immediate admission of such doubts. As I have already demonstrated, there is always the middle way that is excluded by choice. The whole truth always contain such middle way that can be realized through self-inquiry.} It is to accommodate such doubts that I have mentioned ``verbal expressions'' while explaining my view of definition. Nevertheless, the goal is to investigate the extent to which such doubts can be reduced by converting the verbal expressions to physico-mathematical ones. Therefore, from the logician's perspective my view of definition inherits the presumption  that there can be definitions which may not reach a logical completion i.e. even after the construction of a definition from the stated premises there can be an  element of doubt. {\color{black} Kant would prefer the word ``exposition'' instead of the word ``definition'' as far as my view regarding the meaning of ``definition'' is concerned:\vspace{0.1cm} 
	
	{\small ``{\it Instead of the term, definition, I prefer to use the term, exposition, as being a more guarded term, which the critic can accept as being up to a certain point valid, \underline{though still entertaining doubts} as to the completeness of the analysis.}''}\vspace{0.1cm}
	
	 -- see page no. 144, Vol. 1 of ref.\cite{bragg}.}

\subsection{Comparison with the physicist's view}\label{physdef}
My view regarding the meaning of definition is very much in conflict with the operational view generally held by the physicist. The physicist's view is ignorant of the necessity of clear statements about the founding premises based on which a logical construction of some concept can be expressed as a definition. As a result, the operational definitions become plagued with  inconsistent expressions i.e. what is written in terms of verbal language as a manifestation of the immediate experience of the experimenter or observer, drastically differs from the corresponding physico-mathematical counterparts. The following example may   clarify such objections.


Newton's Definition I  in ref.\cite{principia1} can be quoted as follows:\vspace{0.1cm} 

{\small  ``{\it The quantity of matter  is the measure of the same, arising from its density and bulk conjointly.}''}\vspace{0.1cm}

 Certainly, ``quantity of matter'', ``density'' and ``volume'' are words which do not stand for number, otherwise the use of such words would have been unnecessary. Hence, doing arithmetic operations with those words do not carry any meaning unless such meanings are separately assigned to those words through separate propositions outside the realm of arithmetic. Such clarifications are absent even in a modern textbook. For example, on page no. 1 of ref.\cite{fom1}), the authors begin with the following declaration: \vspace{0.1cm}

{\small ``{\it When measuring some attribute of a class of objects or events, we associate numbers (...) with the objects in such a way that the properties of the attribute are faithfully represented as numerical properties.}''}\vspace{0.1cm}

 The doubt arises that if attributes are numbers then, given a number, there can be various attributes. But the expression of any number only carries a sense of that number itself defined in the realm of arithmetic and not multiple senses of different attributes i.e. the number ``three'' is not an attribute like ``length'', ``mass'', ``volume'', etc.\footnote{This is similar to the objection raised by Frege in the opening paragraph of ref.\cite{frege1} about the sentence: ``{\it The number one is a thing}''.}. Otherwise, writing the word ``three'' would have sufficed to mean ``length'', ``mass'', ``volume'', which is at the highest level of absurdity even in our experience of daily life.

 Therefore, there is a clear necessity of an explanation of the sense in which the words like ``length'', ``mass'', ``volume'', etc. are associated with numbers. Such necessity can be fulfilled by stating out the premises and then defining a concept on the basis of those premises, being as less vague as possible.  My view of definition aims at fulfilling such necessities at the elementary level of physics. I shall discuss shortly how basic fallacies plague the definition of electric field in physics
 as a result of  holding fast to a complete operational view of ``definition''.


\section{Quantity, physical dimension and number}\label{qpdn}
The  definitions of derived physical dimensions are taken for granted in physics without any clarification regarding how such non-mathematical symbols, like $[M],[L]$, etc., can be written as if they {\it were} numbers. At least for me, this appears to be an unreasonable way to find reasonable explanations of observed phenomena because the premises need to be stated with clarity when it comes to reason. In this regard, it is also necessary to clarify the distinction between physical dimension and quantity because neither of them are numbers by themselves, nor they are identical concepts\footnote{For example, some quantity of rice is not some number by itself; but it is expressed as number while compared with another  quantity of same physical dimension -- it might be mass dimension, volume dimension, etc.}.  I provide such definitions in the following subsections.

\subsection{Quantity{: dependent origination and self-inquiry}}\label{quantity}
 To define ``quantity'', I consider ``number'' and ``physical dimension'' as the premises\footnote{I make no attempt to define ``number'' (unlike the logician e.g. see ref.\cite{frege1}). 
 	For me, the thought of ``one'' comes from the realization of the self and the non-self. Even a child without any bookish knowledge has the realization of the self in terms of ``one'' and the non-self, which can be either the ``relative zero'' as the environment or the ``two'' as the other person/mother or object (of sensation). Without such realization the child should not cry while out of the mother's womb. Indeed, the child's kick, while still inside the mother's womb, is a prior indication of the realization of the self and the non-self. Otherwise, how can the self kick the non-self.  May be I discuss more on this on a separate occasion, but I believe I could convey my point to the concerned reader.}. The definition goes as follows. 
 
Suppose I represent different quantities of charge by the symbols $q_1,q_2$, etc. Then,
\begin{eqnarray}
q_i:=n_{q_i}[Q]\ni n_{q_i}>0,\label{chargedef}
\end{eqnarray}
where the lower index ``$i$'' on the left hand side is the label to denote different quantities of the same physical dimension and the symbols ``$:=$'', ``$\ni$''stand for ``defined as'', ``such that'' respectively. ``$n$'' stands for the concept of ``number''\footnote{It is straightforward to note that the phrase ``the concept of number'' itself is constructed out of {\it four} words (also countable alphabets). Therefore, the concept of ``number'' is already being used to write the expression ``the concept of number''. I leave it to the reader for a judgment of whether to consider this as an unavoidable self-referencing process or a circular fallacy.}, ``$n$'' carries the lower index ``$q_i$'' to denote which quantity {\it the} number is defining and ``$[Q]$'' denotes the associated type of physical dimension which defines the ``type'' of quantity. 
The same definition goes for other {\it types} of  quantities e.g. mass $m_i:=n_{m_i}[M]\ni n_{m_i}>0$, length $l_i:=n_{l_i}[L]\ni n_{l_i}>0$, time $t_i:=n_{t_i}[T]\ni n_{t_i}>0$, etc. 

{\color{black} Now, I discuss some doubts through self-inquiry and refine the definition (\ref{chargedef}) as follows. 
An immediate objection against the definition (\ref{chargedef}) may be that, ``the definition contains a circular fallacy because, to define `$q_i$', `$q_i$' has been used as the lower index of `$n$' on the right hand side.'' Such objection arises from an expectation of a complete deduction from the specified premises which are considered as independent and fundamental truths rather than dependent and conventional truths. As it appears to me, neither a complete deduction is  possible nor a complete induction is possible. Hence, the circle is only apparent. To elaborate such view, I pose the following question:~ 
{\it Can ``number'' be conceptualized and explained without any reference to ``quantity'' and vice versa?}~ This question can not be answered completely as either ``yes'' or ``no'', which I explain in what follows. 

In physics, examples of ``quantity'' are ``mass'', ``charge'', ``length'', etc., which are different aspects associated with any ``object'' and this is why the modern physicist writes ``mass of an object'', ``charge of an object'', ``length of an object'', etc. Now, while writing such phrases, there are two points to be noted. Firstly, the words ``mass'', ``charge'', ``length'', etc. do not have any independent meaning unless associated with the word ``an object'', but ``an object'' is none of ``mass'', ``charge'', ``length'', etc. Secondly, the word ``object'' unavoidably gets associated with the word ``an'' or ``one''\footnote{Certainly one can use the words ``the object''. However, this indicates the observer who identifies ``the object'' has prior experience of ``an object'' with which he identifies the current experience.}. Therefore, the concepts of ``object'',``one'' and  ``quantity'' (like ``mass'', ``charge'', ``length'', etc.) originate in a mutually dependent fashion, but these concepts are different from each other {\color{black}-- this exemplifies dependent origination.} None of the concepts can be explained completely independent of each other, although each of the concepts are required to be considered as different from each other in order to express the thoughts and write physics. That is, the truth of each categorical words is empty of itself and rather can be understood in relation to the others, which is the essence of emptiness.

 Therefore, ``quantity'' and ``number'' can not be conceptualized and explained completely independent of each other, although the two are different concepts. This is why I mentioned that the above (italicized) question can not be answered completely as either ``yes'' or ``no''. Both the concepts arise in association with the concept of ``object'' which is neither ``quantity'' nor ``number''  i.e. the object holds the essence of the middle way in relation to which arise the concepts of  both ``quantity'' and ``number''.

Thus, the person with a complete answer ``yes'' to the above question\footnote{For example, Peano's arithmetic is founded on such attitude \cite{peano}, which is in contrast to the object oriented foundations of arithmetic discussed by Bhaskaracharya \cite{lilavati}.}, holds the responsibility to explain the concept of ``number'' and arithmetic operations {\it without} referring to objects i.e. he can not use expressions like ``a table'', ``three chairs'', ``a handful of grains'', etc. In other words, the concerned person should be able to explain the meaning of the words ``one'', ``two'', etc. and of the numerals ``1'', ``2'', etc. without saying ``one object'', ``two objects'', etc. Certainly, any sort of such explanation becomes inapplicable for practical purpose where the concept of ``object'' is involved.  On the other hand, the person with a complete answer ``no'' holds the responsibility of explaining the use of   words like ``mass'', ``charge'', ``length'', etc. to explain different types of ``quantity'' instead of just using the concept of ``number'' to represent all types. That is, if ``quantity'' and ``number'' are completely dependent\footnote{``Complete dependence'' can be thought of as being ``identical''. In this case, it would mean that the concept of ``number'' becomes identical to that of ``quantity'' and hence, all types of ``quantity''.} concepts then there should not be any necessity of using so many words such as ``mass'', ``charge'', ``length'', etc. as all these concepts converge to being completely dependent on the concept of ``number''. Above that the responsibility of explaining the role of physical dimension remains there too.  Certainly, such stance seems not to make sense as far as physics is concerned. In a nutshell, the person must go through self-inquiry in order to realize that the answer given in either ``yes'' or ``no'' does not grasp the whole truth.

Thus, to summarize, ``number'' is conceptualized in association with that of ``object''. Hence, ``quantity'', like ``mass'', ``length'', etc., being aspects of  ``an object'', is a concept that is interwoven with the concept of ``number'' and can not be explained completely independent of each other. However, certainly ``quantity'' is not completely dependent on ``number'' which can be understood by the fact that if I omit the symbol $[Q]$ in (\ref{chargedef}), then there is a problem in explaining the difference between any ``quantity'' (e.g. ``length'') and ``number'' and also among different types of ``quantity'' like ``mass'', ``charge'', ``length'', etc. This explains why there is no circular fallacy in (\ref{chargedef}), although it can be realized only through self-inquiry}.

\subsubsection{Avoiding a conflict {and removing the logico-linguistic fallacy of ``zero quantity''} }\label{secconflict}
I have deliberately left out the possibility of $n_{q_i}=0$ because it provides an impression that $q_i$ is non-existent. This conflicts with the premise (i.e. the quantity exists) which needs to hold in the first place because, if there is no quantity (charge) to begin with, what then is getting defined in the definition (\ref{chargedef})\footnote{In this regard, I find the following analogy very useful. A teacher is teaching addition and subtraction to her students. She draws pictures of three apples on the board. Then, she erases one picture and asks, ``How many apples are there?'' The students answer: ``3 apples - 1 apple = 2 apples.'' Then she erases the remaining pictures and asks the same question. The students answer: ``2 apples - 2 apples = 0 apple.'' Now, one of the students was sleeping and now he has woken up when the teacher was asking her last question. Also, he does not see any picture on the board. He gets confused by his teacher's question  and expresses his confusion as follows: ``I do not see any drawing on the board. {\it How can you know what is not there?}'' The same concern was also raised by Frege as he wrote on page no. 11 of ref.\cite{frege1}: ``{\it ..... for up to now no one.... has ever seen or touched 0 pebbles.}'' ``Zero quantity'' is a meaningless statement because if there is ``no quantity'' then how can there be ``quantity'' at all so that the observer can refer to that ``quantity'' as ``zero quantity''. To be referable, some ``quantity'' needs to be sensible or perceptible. To be sensible or perceptible the ``quantity'' needs to exist in the first place and if it exists then how can one explain the meaning of ``zero quantity''. If one explains ``zero quantity'' as ``non-existent quantity'' then nothing has been sensed or perceived at all. Then the student's paradox arises.}.

This can be further clarified in the context of comparison between two quantities e.g. charges $q_1:=n_{q_1}[Q]$ and $q_2:=n_{q_2}[Q]$. Since $n_{q_1}\neq 0\neq n_{q_2}$, then a relation between $q_1$ and $q_2$ can be expressed as follows:
\begin{eqnarray}
	q_2=n_{q_2}[Q]=\frac{n_{q_2}}{n_{q_1}}~n_{q_1}[Q]=n_{21}q_1~\ni n_{21}=\frac{n_2}{n_1}.
\end{eqnarray}
Therefore, for this particular physical dimension $[Q]$
\begin{eqnarray}
&&n_{21}>1\iff q_2>q_1\qquad\[\text{in words: ``$q_2$ is greater (bigger, larger) than $q_1$'' }\]\\
&&n_{21}<1\iff q_2<q_1\qquad\[\text{in words: ``$q_2$ is less (smaller) than $q_1$'' }\]\\
&&n_{21}=1\iff q_2=q_1\qquad\[\text{in words: ``$q_2$ is equal to $q_1$'' }\].
\end{eqnarray}
Further, let me note the following:
\begin{eqnarray}
&&n_{21}\lll 1\iff q_2\lll q_1\nonumber\\
&&~~~~~~~~~~~~~ \[\text{in words ``$q_2$ is infinitesimally (very very, extremely) smaller than $q_1$''}\].~
\end{eqnarray} 
If $n_{21}=0$ {\it were} allowed (which is only possible if $n_{q_i}=0$ {\it were} allowed), then it leads to the expression ``$q_2=0 q_1$''. The corresponding verbal expression should mention both the charges. However, I can not find such a meaningful verbal expression. A possible verbal expression may be ``$q_2$ does not exist (vanishes) compared to $q_1$'', which is however meaningless and grammatically erroneous. {\color{black} I consider this as a logico-linguistic fallacy  that is clearly manifest during the process of conversion of verbal statements to corresponding physico-mathematical expressions. The fallacy is simply removed by excluding the use of an exact zero in the definition of quantities. Nevertheless, such a verbal expression can still be forcefully considered to be meaningful, which I consider as the imposition of dogma that is against the scholarly spirit of science.}

\subsubsection{Existence of quantity {and the essence of emptiness}}\label{existencequantity}
 The definition of quantity in (\ref{chargedef}) involves {\it only} the reference to the particular quantity that is getting defined i.e. $i=1$ implies the definition of $q_1$, $i=2$ implies the definition of $q_2$ and so on. That is, I have considered  independent existence of quantity while defining the same. On the other hand, comparison of some quantity $q_1$ with another quantity $q_2$ of the same physical dimension is necessary in order to give meaning to measurement, which provides the practical or empirical knowledge of the existence of some quantity. 

 In this sense, the number $(n_{q_i})$ that appears in the definition (\ref{chargedef}) can not be known in terms of measurement, but can only be assumed for theoretical analysis. I call such a number as {\it absolute measure}. What can be known in terms of measurement is only the number $n_{21}$. I call such a number as {\it relative measure}. That is, the practical (empirical, knowable) existence of some quantity is relational i.e. dependent on another quantity. 
  Therefore, notion of `existence of quantity' is neither completely independent, nor it is completely dependent. While comparing two quantities, both need to be defined independently from each other, but the comparison itself is relational.  However, in either way, an ``exact zero'' leads to a conflict.
  
  {\color{black}From a careful examination the question arises whether the definition of quantity in (\ref{chargedef}) is really given in  isolation. Certainly it is not, because I  observe the physical phenomena and interpret my experience in terms of physico-mathematical expressions. Thus, the definition of quantity in (\ref{chargedef}) is given in relation to the ``I''. The role of the ``I'' can be realized through self-inquiry i.e. ``I write the definition (\ref{chargedef}).'' The involvement of the conflicting categories like ``absolute'', ``relative'' provides the essence of the middle-way and dependent origination in the definition of quantity. As a whole, the existence of quantity is relational and hence, empty of any isolated truth -- this is the essence of emptiness. 
  

 Newton's definition of quantity was written as ``measure of the same'', for example, in Definition I at the beginning of ref.\cite{principia1} and on subsequent occasions. There was no clarification whether such measure is hypothetical (like the absolute measure) or an experimental measure obtained through measurement (like the relative measure)\footnote{Unfortunately, even the most modern textbooks of ``measure'' theory does not provide such basic clarifications e.g. see ref.\cite{measure}}. If it were the former, certainly Newton's claim to provide explanation for his experiments were false (in the paragraph that follows Definition I). If it were the later, then Newton's statement does not reflect a relation between two different quantities so that a relative measure could be realized. In this later case, a proper statement should obviously have reference to measuring units with respect to which the measurements were performed. Unfortunately, a study of the later sections of ref.\cite{principia1} only reveals statements like ``velocity= 2'' where there is no reference to any measuring unit. Defending such ill-written statements in the name of being ``operational'' \cite{bridgman}, while doing physics, seems to me as being immoral. This is certainly not a question of modern and ancient physics, but a question of how well reasoned the statements of physics are.
 
 Nevertheless, in light of the definition given in (\ref{chargedef}), it appears to me that Newton's ``measure of the same'' is indicative of the absolute measure $n_{q_i}$, which can be explained through the following map: ``$n$'' stands for ``measure'', ``$q_i$'' stands for ``the same'' because the quantity $q_i$ is getting defined and ``$q_i$ appears in the lower index'' denotes the conjunction ``of'' i.e. an association that can not be dissociated. }    
  

\subsection{Physical dimension $[Q]${: neither number nor not number}}\label{physicaldimension}
Since arithmetic operations are only meaningful for numbers ($n_{q_i}$), application of the same for  quantities $(q_i)$ (e.g. $q_1+q_2, q_1\times q_2, q_1/q_2,$ etc.), necessitates a separate declaration regarding the treatment of $[Q]$ as number (or not) itself  in the form of a postulate. It may appear, at a first glance, extremely naive to raise such a concern at all. However, if a bit of care is taken in the way of reasoning along such line of thought, the foundations of physics appear to become untenable from the logician's perspective. I explain as follows.\vspace{0.1cm}

{\bf $[Q]$ does not represent number....~:} In  (\ref{chargedef}), I have defined quantity in terms of number and physical dimension. Since there are two expressions involved i.e. `number' and `physical dimension', then the associated concepts are distinct i.e. $[Q]$ does not represent number. Unlike ``$1+1=2$'' in arithmetic, ``$[Q]+[Q]$'' does not represent counting, even if such written expression is considered meaningful at all. On similar grounds, unlike ``$2\times 2=4, 2\times 3=6$'' in arithmetic, ``$[Q][Q],[M][L]$'' do not imply multiplication unless these expressions are postulated separately as definitions outside the realm of arithmetic.

{\bf .... but, $[Q]$ used as number:} Now, let me consider the following physico-mathematical expressions involving two quantities of charge $q_1$ and $q_2$:
\begin{eqnarray}
&& q':=	q_1+q_2 \iff n_{q_1}[Q]+n_{q_2}[Q] \iff (n_{q_1}+n_{q_2})[Q]\iff n_{q'}[Q]~\ni n_{q'}:=n_{q_1}+n_{q_2},\qquad
 \end{eqnarray}
 $[Q]$ is just written symbolically as if it {\it were} a non-zero number. Further, I consider the following operations with quantities: 
 \begin{eqnarray}
&&q_1\times q_2=n_{q_1}[Q]\times n_{q_2}[Q]=(n_{q_1}\times n_{q_2})[Q^2]\label{q1q2}.
 \end{eqnarray}
 To write down the above expressions there needs to be the following definition:
 \begin{eqnarray}
 	[Q^2]:=[Q][Q]\label{defdim}.
 \end{eqnarray}
This definition justifies such way of writing by postulate. Otherwise there is no meaning of such written expressions because $[Q]$ is not number and such operations are not pre-explained by any stated axioms i.e. ``$[Q][Q]$'' does not mean ``$[Q]+[Q]+[Q]+\cdots [Q]$ times'' because $[Q]$ is not number. Here, I have considered ``$[Q^2]$'' and ``$[Q]^2$'' as the equivalent way to express the same thought. Similar explanation is needed for $m_1\times m_2$. 

Without such explanation, the definition of derived physical dimension in (\ref{derdim}), dimensional analysis, and hence physics, becomes meaningless. 

\subsection{Physics and middle way{: addition of two quantities (an example)}}\label{physicsmiddleway}
Considering the discussion of subsection(\ref{physicaldimension}), there arises a necessity to explain the apparent logical dilemma regarding the status of physical dimension $[Q]$ if one regards physics to be a subject that is meant for logical explanation of observed phenomena. This is because the founding premises of physics do not seem to obey formal logic that is based on the Aristotelian attitude of considering {\it absolute} truth or falsehood of some statement \cite{aristotlepa}. 

Here, the concerned statement is
\begin{eqnarray}
S: ~[Q]~\text{is not number}.\nn
\end{eqnarray} 
It appears that the validity of the statement depends on the context.  Therefore,  either physics is illogical if the law of excluded middle is accepted, or, physics is beyond formal logic requiring the abolition of the principle of excluded middle. I clarify the situation as follows.

While defining quantity in (\ref{chargedef}), I need two distinct concepts, namely, those of number (``$n$'') and physical dimension (``$[Q]$''). Otherwise, it is impossible to express the physicalness of quantity that is conceived in thought. Therefore, the statement $S$ is true while quantity is defined. 

Now, to give meaning to the arithmetical operations with quantities like that of numbers, $[Q]$ needs to be considered as number. Even before that, the definition of any derived physical dimension in (\ref{derdim}) and dimensional analysis require the treatment of physical dimensions as numbers too. Therefore, $[Q]$, or physical dimension in general, is treated as number while arithmetic operations are performed with quantities, derived physical dimensions are defined and dimensional analysis is performed.

 If $S$ is {\it absolutely}\footnote{``Absolutely'', ``unconditionally'', ``completely'' -- all three words can be used. } true then, only the definition of quantity is possible, but any arithmetic operation with quantities, definition of derived physical dimensions and dimensional analysis become meaningless. If $S$ is {\it absolutely} false, then a distinction between quantity and number becomes impossible which in turn renders the terms ``mass'', ``charge'', ``length'', ``time'', etc. meaningless as they all become devoid of any expression of physicality. Therefore, neither $S$ nor $\neg S$ is {\it absolutely} true or {\it absolutely} false as the validity of $S$ or $\neg S$ depends on the context in which it is used. Then, I can also conclude that any physico-mathematical expression (i.e. involving quantities and numbers) is meaningful if and only if both $S$ and $\neg S$ are contextually valid. I may clarify this statement by explaining the role of $S$ and $\neg S$ in  writing  $q=q_1+q_2$ as follows
 \begin{eqnarray}
 	\underbrace{\underbracket{~~q~~}_{\text{definable iff }  S}=\underbracket{~~q_1~}_{\text{definable iff }  S}+\underbracket{~~q_2~}_{\text{definable iff }  S}}_{\text{meaningful iff } \neg S}.
 \end{eqnarray}

 Thus, it appears that  unlike the two options $S$ and $\neg S$ in Aristotelian sense, ``$P:=$ neither $S$ nor $\neg S$'' and ``$\neg P:=$ both $S$ and $\neg S$'' are {\it not} like ``third'' and ``fourth'' formal options or logical values. Rather $P$ and $\neg P$ provide the informal ground  for {\it operation} with the associated concepts through the use of $S$ and $\neg S$ as per requirement.

 \section{Unit and unity}\label{unitunity}
  Now, it is important to clarify the difference between {\it unit} and {\it unity}\footnote{Necessity of such clarification was also pointed out earlier by Frege \cite{fregeoneunity}.}. When I consider some quantity of charge, say $U$, as the chosen standard of measurement and call it unit charge, then I write 
  \begin{eqnarray}
     U:=n_U[Q]\label{defunit}
  \end{eqnarray}
  i.e. unit carries the physical dimension. But, unity is just the name of the number ``one'', represented by the Hindu(-Arabic) numeral ``$1$''. Therefore, the concept of unity arises as a comparison of two quantities of charges if they are equal e.g. if I say that there is one unit of charge $q_1$, then I write $q_1=1U$ where both $q_1$ and $U$ are already defined in accord with (\ref{chargedef})\footnote{ In usual sentences like ``This is a table'',  ``I have an apple'', etc. it appears that no comparison between two objects is taking place and hence, the use of ``a'' and  ``an'', which correspond to the number ``one'', seems not to be justified by the above clarification. However, the person who makes such statement presumably objectifies himself and makes such comparison with regard such objectified self. {\color{black} Such explanation can be realized through self-inquiry only. Also, such realization can only dissolve the irony in the statement where I have written ``two'' to explain ``one'' or ``unity'' i.e. ``one''-ness and ``two''-ness originate dependently and can not be explained in isolation. This is an essence of dependent origination. }}. Some quantity can only be expressed as a multiple of the chosen unit i.e. $q=N_q U$ where $N_q=1,2,3,\cdots$ and some quantity smaller than the chosen unit is considered as error. If units smaller than the chosen unit can be made accessible then the error can be reduced. However, there is always a left over quantity that is immeasurable. Therefore, on practical grounds $q_1=n_1 U\ni n_1:= (N_1+\epsilon_1),~ 0<\epsilon_1<1$ where $\epsilon_1$ is  the leftover while $q_1$ is measured in terms of the unit $U$. 
  
   \subsection{A conflict of attitude: theorist vs experimentalist}
    I may note that my view regarding units and measurement is partially in conflict with that of Einstein's view regarding  the same as he wrote in the first footnote on page 4 of ref.\cite{einphil2}:\vspace{0.1cm}
    
    {\small  ``{\it Here we have assumed that there is nothing left over, i.e. that the measurement gives a whole number. This difficulty is got over by the use of divided measuring-rods, the introduction of which does not demand any fundamentally new method.}''}\vspace{0.1cm} 
     
      According to Einstein, the error in measurement can be removed by refining the unit i.e.  $\epsilon_1=0$ can be achieved in practice.\footnote{For some discussion and comments on Russell's view on such issues,  one may consult Appendix \ref{appb}.}  Since Newton, such view has been continued to be adopted throughout the physics literature when theories are written down \cite{essay}. 
      
      However, such view is in straightforward disagreement with everyday experience in general for any common man and in particular with the view of an expert experimentalist like Bridgman as he wrote on page no. 33 of ref.\cite{logicphysics}:\vspace{0.1cm} 
      
      {\small ``{\it .... that all results of measurement are only approximate. That such is true is evident after the most superficial
   	examination of any measuring process; any statement about numerical relations between measured quantities must always be subject to the     	qualification that the relation is valid only within limits.}''}\vspace{0.1cm}

    In spite of such admission by the experimentalist, it is the attitude of the theorist that is manifested in writing. Such examples plague the literature of physics beginning from the time of Newton \cite{principia1,principia2} till date. It is unfortunate that the experimentalist bases the analyses on the theories written by the theorists in a language that does not express what the experimentalist experiences. Such mismatch between  experience and written expression is the founding stone on which the dogma of physics has been built upon.

    \subsection{Experience vs written expressions}
    The mismatch between experience and written expressions is evident since the formal beginning of physics.  For example, on page no. 2 of ref.\cite{principia1}, in the last line of the paragraph after Definition I, Newton declared that his intention was to explain experimental results:\vspace{0.1cm} 
    
    {\small ``{\it I have found by experiments on pendulums, very accurately made,...}''.}\vspace{0.1cm}

    Notably, he did verbally mention ``very accurately'' and not ``exactly''. And as far as my knowledge of English language is concerned, ``very accurately'' means ``with negligible error'' and ``exactly'' means ``with no error''.

    However, when it comes to written physico-mathematical expressions, the story looks different. For example,  on page no. 18 of ref.\cite{principia1}, in Corollary III after the statement of Law III, Newton wrote:\vspace{0.1cm} 
    
    {\small ``{\it Thus, if a spherical body A is 3 times greater than the spherical body B, and has a velocity = 2, and B follows in the same direction with a velocity = 10, then the motion of A : motion of B = 6 : 10. Suppose, then, their motions to be of 6 parts and of 10 parts, and the sum will be 16 parts.}''}\vspace{0.1cm} 
    
    The exact numbers and the exact equalities do not justify the  experimentalist's experience of approximate measurement. Newton's written physico-mathematical expressions do not justify his words ``very accurately done'' and rather justify ``exactly done'' measurements.  Such written statements plague the literature of physics and the attitude only reverberates through the words of Einstein that I quoted earlier in the last subsection.

    Further, there are theoretically inconsistent expressions. For example, the expressions like ``velocity= 2'', etc. are devoid of the units.  One can justify this on the ground that the ratio of quantities was being considered by Newton and therefore the units did not matter. However, that does not justify that written expressions like  ``velocity=2'' are meaningful because the left hand side expresses some quantity (i.e. associated with physical dimension) and right hand side expresses number (i.e. not associated with physical dimension). Similar situation also arises when squared line element is equated to number in relativity viz. $(\Delta s)^2=0, (\Delta s)^2=-1$, with the physical dimension of $[L]^2$ on the left hand side and number on the right hand side (see e.g. refs.\cite{eingr,landau}). In a nutshell, physics is written in terms of such inconsistent expressions.

    
    \subsection{Logical vs Operational: morality of expression {-- do ``you'' see the dot or the cross-wire?}}\label{doyousee}
      Any excuse from the {\it modern} physicist in the disguise of having more sophisticated theories in the present days is insufficient to cover up the inconsistent expressions in terms of which physics is written. Solving some problem, that was created due to inconsistent expressions, with added hypotheses or postulates containing sophisticated levels of abstraction, does not explain why such misleading expressions were written in the first place. Although an experimental observation must have the last say in the success of any theory in physics, but there is no denying that the experimental results need interpretations in terms of the expressions written in the theory. 
      And if both the theorist and the experimentalist rely on the same inconsistent expressions of experience then the experimentalist's observation would certainly not defy the theory as far as the basic elements of expression are concerned. 
      
      A straightforward example is that of the cross-wire of an eyepiece of a microscope. The existence of the cross-wire, which is realized by the experimenter due to its thickness, is an irremovable error in the measurement. Whatever be the resolution ability of the microscope, if the cross-wire is not there then measurement is impossible and if the cross-wire is there then it itself becomes the irremovable error. The question is whether the experimenter takes account of this irremovable error while noting down the numbers that the measurement yields. The answer is certainly negative. So does the theorist who assigns exact zero to a dot put on the paper with a pencil \cite{essay}. It is a direct implication of Cartesian philosophy of representing real number system by drawing a line, which depends on the argument that even if a dot can be seen with naked eye due to its finite spread, one can call it to have ``exactly zero extension''. {\color{black} Although such is the usual practice in modern days, however it appears to me at least, that  Descartes was very much aware that such pictorial representation does not hold the exact truth of numbers and rather provides a useful but approximate mode of representation. Descartes' view becomes apparent from his statement on page no. 2 of ref.\cite{descartes}:\vspace{0.1cm} 
      	
      	{\small ``{\it ..... taking one line which I shall call unity to relate it \underline{as closely as possible} to numbers.}'' }\vspace{0.1cm}
      	
      	Here, Descartes did hesitate to write that a line, as a whole, exactly represent the number unity. Otherwise, he would have written ``exactly'' instead of ``as closely as possible''.} And this brings back the question whether exact zero can be represented at all, considering that any method of representation needs a physical act: If I can see the dot how can I call it ``zero'' and if there is no visible dot then what can I refer to as having ``zero extension''?

      So, the operational approach of physics, as often  advocated by the physicist \cite{logicphysics}, if considered to be a self-complete approach towards scientific inquiry, then such a method of inquiry is primarily based on such elementary inconsistent expression of experience. Treating such expressions of experience as ``consistent'' and ``exact'' in disguise of being ``operational'' can only be deemed as an untruthful approach to scientific queries if science is considered to be meant for reasonable explanations of physical experiences in terms of written physico-mathematical expressions. Since the physicist's operational approach is based on practicality of the expressions, and certainly such inconsistent expressions have been capable of providing very useful  advancement in technology, the logician should not stand a chance to question the logical structure of the expressions and reasoning. The issue then boils down to morality of the expressions.
      In such a scenario the logician should take now an operational approach and ask the physicist to perform an experiment with a faulty microscope that does not have a cross-wire. Certainly the experimentalist would take a step backward, but the theorist may argue, ``One can not write down everything that is experienced''. Then, the logician should ask, ``If that is the case, then how can ``you'' write down a theory of everything\cite{hawking} or even dream of a final theory\cite{weinberg} or provide a complete guide to reality\cite{roadtoreality}?''
      
      {\color{black} Certainly the expression of the perception of the dot or the cross-wire becomes a matter of self-inquiry like the scenario of the closed line in subsection (\ref{closedline}). Such doubts, if categorized as `philosophical', then certainly philosophy of physics is devoid such self-inquiry. The reader may consult ref.\cite{companion} for some further remarks.}

      


\section{The physicist's definition of field: analysis with reasoning}\label{physdeffield}

For the physicist it suffices to have just an operational character of some concept and its definition in terms of other concepts which are also operational in nature \cite{bridgman}.  As far as the concept of field is concerned, the definition is no exception. Therefore, considering this as an example, I intend to bring out the elementary fallacies that can plague the foundations of a subject if such complete operational character of definition is adopted and basic logical necessity for consistent expressions of experience is simply ignored.  The necessity of such an endeavour arises at a stage while one intends to write down some expression corresponding to the  thought of field that has been conceived by observing some physical phenomena. Therefore, the associated queries are sequenced prior to more complex derivations of thought such as ``theory of fields'' and consequent further categorizations such as ``classical''\cite{landau} and ``quantum''\cite{coleman}.


 \subsection{Definition of electric field}  
 The definition of electric field, as considered by the physicist, involves the Coulomb's law as a premise and associated with some further explanations as I state below.
 
 Coulomb's law is stated as follows:-    The magnitude of the force $(F)$ between two objects having charges $(q_1,q_2)$, which are at rest with respect to each other, is proportional to the product of the two charges and inversely proportional to the square of the distance between those two point charges. This statement is expressed as 
\begin{eqnarray}
F=k \frac{q_1q_2}{r^2}\label{claw}
\end{eqnarray}
where $k$ is a proportionality constant that needs to be fixed by experiment. $r$ is the geometric expression of  distance between the two charges considering them as points and it is defined from the axioms of geometry.


From expression (\ref{claw}), the electric field due to $q_2$ (source charge), is {\it defined} as
\begin{eqnarray}
E(q_2):=\lim_{q_1\to 0}\frac{F}{q_1}=k\frac{q_2}{r^2}.\label{q1to0}
\end{eqnarray}
$q_1$ is called the test charge (e.g. see page no. 16 of ref. \cite{purcell}, page no. 29 of ref.\cite{schwartz}, page no. 25 of ref.\cite{rmc}).

There are two issues that I intend to discuss here. At first I comment on the problem associated with the symbolic statement ``$\lim_{q_1\to 0}$'' in general. Then I discuss the problems of the usual interpretations of the statement ``$\lim_{q_1\to 0}$'', with regard to the definition of electric field, which are available explicitly or implicitly in the existing literature.

\subsection{Meaning of ``$\lim_{q_1\to 0}$'' in general}
It is quite straightforward to argue that the symbolic statement ``$\lim_{q_1\to 0}$'' is meaningless\footnote{I may clarify that there is no conflict with mathematics where $q_1$ represents a number and not quantity. However, with only mathematics there can be no application for practical purpose. One needs to introduce physical dimension so as to express the thought of physicality.}. $q_1$ is not a number by itself, but represents a quantity (charge) defined in accord with (\ref{chargedef}). If it is compared with another  quantity (charge), say $C$ (also defined in accord with (\ref{chargedef})), then such comparison is expressed by a relation like $q_1=\epsilon_1C$ where $\epsilon_1$ is a number. Now, if one writes ``$\epsilon_1\to 0$'', it seems to be a justified statement as it represents the relation between two numbers viz. $\epsilon_1$ and $0$. This automatically justifies why ``$q_1\to0$'' does not make sense at all because on the left hand side there is a quantity $(q_1)$ and on the right hand side there is a number $(0)$ \footnote{Such statements plague the foundations of calculus  \cite{comment}.}. Therefore, the symbolic statement ``$\lim_{q_1\to 0}$'' should be replaced by one like ``$\lim_{\epsilon_1\to 0}$''. But, there is a further issue in this particular way of writing. Generally in mathematics, if there is an expression like $y=\epsilon x$, where  $x$ is a real variable (i.e. number), $y$ is its function and $\epsilon$ is a chosen real parameter  then one writes $\lim_{\epsilon\to0}y=\lim_{\epsilon\to0}\epsilon x=0$ i.e. {\it exactly} $\epsilon=0$ is put in the expression. But, if I write 
\begin{eqnarray}
\lim_{\epsilon_1\to0}q_1=\lim_{\epsilon_1\to0}\epsilon_1C=0C,\label{limq}
\end{eqnarray}
then there is a problem in the  interpretation. The sequence of expressions in (\ref{limq}) provide an impression that $q_1$ has vanished or ceased to exist \footnote{I find the situation quite akin to Berkeley's ``ghosts of departed quantities'' in ref.\cite{analyst} - a critical remark meant for Newton's ``evanescent'' quantities that founded the basis of calculus\cite{principia1}.}. However, as I have already clarified, $q_1$ and $C$ both are quantities and for a comparison to happen, both the quantities need to {\it exist} in the first place. If one of the quantities does not exist (here $q_1$) or I have only  one quantity at my disposal (here $C$), then a situation for comparison does not even arise in the first place. Therefore, the operation meant by the sequence of expressions (\ref{limq}), although looks mathematically legitimate, leads to a contradiction with regard to physical interpretation i.e. the conclusion defies the premise. In order to save the situation, there must be a restriction like $\epsilon_1>0$ i.e. {\it $\epsilon_1$ can be arbitrarily close to zero, but can never be exactly equal to zero.} Rather, to distinguish the situation from mathematical scenario, it seems more reasonable to write ``$q\ll C \equiv \epsilon_1\ll 1$'' in order to mean ``$q_1$ is extremely\footnote{The word ``extremely'', here and henceforth, can be replaced by ``infinitesimally'' or ``arbitrarily'' in accord with the conventional practice or the convenience of the concerned reader.} small compared to $C$''. Therefore, in what follows, wherever required I shall use a notation ``$\lim_{\epsilon\ll1}$'' instead of the notation ``$\lim_{\epsilon\to 0}$'' so as to avoid any {contradiction}, unless otherwise stated.

\subsection{Interpretations of ``$\lim_{q_1\to 0}$'' while defining electric field}
For the moment, I disregard the above discussed issues with the statement ``$\lim_{q_1\to 0}$''. Rather, in what follows, I discuss the two different kind of explanations which can be given, from an operational point of view, in order to justify the use of the statement ``$\lim_{q_1\to 0}$'' as a part of the definition of electric field \cite{purcell,schwartz}.  I argue that while one of the interpretations is circular in reasoning, the other one is incomplete.  To mention, I shall consider only positive test charge as per usual convention.  

\subsubsection{Circular reasoning}\label{cr} 
The first explanation goes like this\cite{reason}:\vspace{0.1cm}

  {\it The statement ``$\lim_{q_1\to 0}$'' is used to signify  that the field due to the test charge $q_1$, must not distort the field due to $q_2$, or must not affect the body with charge $q_2$ .}
  \vspace{0.1cm}

Certainly, to affect $q_2$, $q_1$ must produce a field $E(q_1)$ and what one  actually assumes in the above way of reasoning is that ``$q_1\to0$''~$\equiv~$``$E(q_1)\to 0$'', or more explicitly
\begin{eqnarray}
E(q_2):=\lim_{E(q_1)\to 0}\frac{F}{q_1}=k\frac{q_2}{r^2}.\label{Eq1to0}
\end{eqnarray}
 Such an explanation is circular in reasoning because, in order to {\it define} the field (due to $q_2$), one {\it explicitly presumes} that the field (due to $q_1$) is already defined. To avoid this circle one may propose to consider another test charge $q_0$ in order to {\it define} the field due to $q_1$ so as to make the presumption justified. However, this would imply that one needs a statement like ``$\lim_{q_0\to 0}$'' for the test charge $q_0$,  which in turn again leads to the same circular reasoning, i.e. to define the field due to $q_1$ one needs to explicitly presume that the field due to $q_0$ is already defined, and provide an explanation like this --     {\it The statement ``$\lim_{q_0\to 0}$'' signifies that the field due to $q_0$ does not distort the field due to $q_1$}. Therefore, one falls in a circular trap.

\subsubsection{ The physicist's immoral objection to circular reasoning} \label{iob}
The physicist may argue that such an apparent circular trap is not a {\it vicious} circle like what Tarski has warned in course of a logically constructed  definition on page no. 31 of ref.\cite{tarskidef}. This is because, although it appears that ``electric field is getting defined on the basis of a presumed definition of  electric field'', but a careful observation shows that ``electric field due to $q_2$ is getting defined based on a presumed definition of electric field due to $q_1$''. Since {\color{black}$q_1$ and $q_2$ are distinct charges}\footnote{\color{black}It is tempting to write ``$q_1\neq q_2$'', which conveys the message to the physicist that ``$q_1$ and $q_2$ are unequal in magnitude''. This is certainly not what I intend to write and hence, I have opted to explain my thought verbally.} then the circle in not exact and the fallacy is not obvious. Moreover, such definition of electric field is operational.

Nonetheless, from the logician's perspective, the use of the word ``electric field'' to define the same, certainly still appears problematic. Indeed, a careful analysis of the expressions only reveals that the main fallacy of the above interpretation of ``$\lim_{q_1\to 0}$'' is the associated immorality. This is because the symbol ``$q_1$'' represents the word ``charge'' and not ``electric field due to some charge''. So, the explanation must not involve the words ``electric field''. Therefore, the fallacy is incorporated in such interpretation while the physicist associates the words ``electric field'' in order to verbally explain the meaning and necessity of the use of  ``$q_1\to0$''.

\subsubsection{ Incomplete reasoning{: an unresolved contradiction and the oil drop experiment}}\label{oildrop} 
  In order to avoid such circular reasoning and immoral interpretation, the physicist can choose to interpret the statement ``$\lim_{q_1\to 0}$'' as follows:\vspace{0.1cm}
\begin{center}  
{\it $q_1$ is extremely small compared to $q_2$. In this way one need not presume $E(q_1)$ to define $E(q_2)$.}
\end{center}
Now, this is again (like the case in the previous subsection) an immoral interpretation because the statement ``$\lim_{q_1\to 0}$'' does not contain any reference to $q_2$. Such a verbal statement about comparative ``smallness'' (or ``largeness'' if interpreted otherwise) should be written physico-mathematically as ``$q_1\ll q_2$'' or equivalently `` $q_1=\epsilon_{12} q_2\ni 0<\epsilon_{12}\ll 1$'', assuming that one can actually find a way to compare the two charges physically without hampering the system itself. 

Thus, the question has now shifted from ``how to define field'' to ``what is the required premise on which the definition of field can be founded, without circular reasoning''. That is, the situation is like this now -- one has already  conceived  the notion of field in thought and there is now a search for some additional proposition, alongside  Coulomb's law,  from which the definition of field can follow without relying on an assumption of the same \footnote{I find the situation like a trial to avoid the vicious circle fallacy of the  statement ``{\it I am lying.}'' (see page 64 of ref.\cite{pm1}).}. 
 
 However, now the question arises that if one needs $q_1\ll q_2$ to define $E(q_2)$, then one needs some $q_0\ll q_1$ to define $E(q_1)$ and then $q'\ll q_0$ and so on, considering that $E(q)$ should be definable for any $q$. Now, an immediate question follows -- is there a smallest charge, say $Q_u$, one needs to consider such that $Q_u$ can be used as the test charge to define the field due to any other (source) charge $q_s\gg Q_u$?

Nevertheless, since Coulomb's law does not come with a restriction on the charge unit, $Q_u$ should obey the law. Hence, the thought of  $E(Q_u)$ has already been conceived when it is considered to obey Coulomb's law  i.e. $Q_u$ exerts a force on the source charge whose magnitude is given by Coulomb's law. Now, in order to define $E(Q_u)$ one needs to assume a smaller charge, say $Q_0<Q_u$ and this violates the assumption of $Q_u$ to be the smallest charge. Hence, this leads to a useless contradiction that  does not serve the purpose of setting the premise on which the field can be defined. 

Therefore, I find this above way of interpretation of the statement ``$\lim_{q_1\to 0}$'', while defining electric field, to be incomplete {\color{black} because it ends up in an unresolved contradiction. Importantly, this contradiction plagues the work of Millikan \cite{millikan}(and Fletcher \cite{fletcher}) on the experimental verification of discrete or elementary charge, known as ``the charge of an electron'',  through the oil drop experiment. Millikan used theoretical calculations in ref.\cite{millikan} to explain the experimental observations, where he used the concept of electric field explicitly -- he called it ``electric field strength'' and denoted it by the symbol ``$F$'' in ref.\cite{millikan}, rather than the symbol ``$E$'' that I use here. Now, for the definition of electric field to hold, and hence to give a meaning to the symbol ``$F$'', Millikan had to consider the following statement, say $M$, to be true:\vspace{0.1cm}
	
	{\it M: Charge can be arbitrarily small. }\vspace{0.1cm}

 Hence, the theory used by Millikan in ref.\cite{millikan} is valid if and only if $M$ is true. But, Millikan concluded from his experiment, based on such theory, that there is a smallest charge, which is equivalent to the falsification of $M$: \vspace{0.1cm}
 
 {\it $\neg$M: Charge can not be arbitrarily small. }\vspace{0.1cm}

 Therefore, ref.\cite{millikan} is a demonstration of a logical contradiction i.e. Millikan's conclusion contradicts the premise on which he built his arguments.}

 \section{{Physico-mathematical analysis of verbal statements}} \label{pmvs}
  
The distinction in the ways of reasoning that I have discussed in the previous section are not even realized in the existing literature and also sometimes the statement ``$\lim_{q_1\to 0}$'' is not explicitly used. However, the verbal interpretations, which are provided along  with the definition of electric field, convey the relevant message. In order to connect the above discussion to the existing literature I comment on the relevant excerpts from two classic texts, namely, those of Maxwell \cite{maxwell} and Jackson \cite{jackson}, where the subtle  details involved in the definition of electric field are available. {\color{black} In due course of this discussion relevance of the conversion of verbal statements to physico-mathematical expressions should be apparent.}

   \subsection{{Comments on Maxwell's definition of electric field}}
 I quote below the relevant statements from page no. 45 of ref.\cite{maxwell} where Maxwell introduced electric field: \vspace{0.1cm}

\begin{singlespace}
	 {\small ``{\it 
  	If an electrified body be placed at any part of the electric field it will be acted on by a force which will depend, in general, on the shape of the body and on its charge, if the body is so highly charged as to produce a sensible disturbance in the previous electrification of the other bodies. But {\color{blue}if the body is very small} and {\color{redish}its charge also very small}, the electrification of the other bodies will {\color{blue}not be sensibly disturbed}, and we may consider the body as indicating by its centre of gravity a certain point of the field. The force acting on the body will then be proportional to its charge, and will be reversed when the charge is reversed. Let $e$ be the charge of the body, and $F$ the force acting on the body in a certain direction, then when {\color{redish}$e$ is very small} $F$ is proportional to $e$, or	
  	\begin{eqnarray}
  	F=Re.
  	\end{eqnarray}
  	where $R$ is a quantity depending on the other bodies in the field. If the {\color{blue}charge $e$ could be made equal to unity} without disturbing the electrification of other bodies we should have $F=R$. We shall call $R$ the Resultant electric force at the given point of the field.}''}
  \end{singlespace}
  \vspace{0.1cm}

  Now, I discuss how Maxwell's  definition of electric field, although does not explicitly use the statement ``$\lim_{q_1\to 0}$'', but contain the kind of ill-reasoned verbal explanations those are akin to what I have discussed.
  \begin{enumerate}
  \item ``{\it If an electrified body 
  	........ electrification of the other bodies.}'':  This statement is circular in reasoning along the line of arguments those I have discussed in CR. 
  
  \item {\color{blue}``{\it if the body is very small}''}:\vspace{0.1cm}
  
  This statement is incomplete because there is no mention of `with respect to  which body' the concerned ``body is very small''. Also, there is no mention of which quantitative aspect is being compared -- is it mass or volume or density or any other aspect? If I assume that Maxwell wanted to mean ``mass'', then there must be some mass (i.e. quantity with mass dimension), say $M$, with respect to which mass of the body, say $m$, is very small. This, physico-mathematically, can be expressed as $m\ll M~ \equiv ~m=\epsilon_m M~\ni\epsilon_m\ll 1$.

  \item {\color{redish}``{\it its charge also very small}''}:\vspace{0.1cm}
  
  Again, this statement is incomplete because there is no mention of `with respect to which charge' the concerned ``charge (is) also very small''. To have such comparison, there must be some charge (i.e. quantity with charge dimension), say $C$, with respect to which the concerned charge, say $q$, is very small. This, physico-mathematically, is expressed as $q\ll C~ \equiv ~q=\epsilon_q C~\ni\epsilon_q\ll 1$.

  \item {\color{blue} ``{\it not be sensibly disturbed}''}:\vspace{0.1cm}
  
  This is an assumption of the presence of non-sensible disturbance i.e. there has to be some force responsible for sensation, say $F_0$, with respect to which the concerned force, say $F$, causing disturbance must be negligible. This, physico-mathematically, can be expressed as  $F=\epsilon_FF_0\ni \epsilon_F\ll1$. 

  \item {\color{redish}``{\it $e$ is very small}''}:\vspace{0.1cm}

    Again, there is no mention of any quantity (charge) with respect to which ``$e$ is very small''. Therefore, such a statement is incomplete and meaningless. It can be completed by assuming that  $e$ is very small compared to $C$ and, physico-mathematically, can be expressed as $e\ll C ~\equiv~ e=\epsilon_eC~\ni \epsilon_e\ll 1$.
    
    \item {\color{blue}``{\it charge $e$ could be made equal to unity}''}:\vspace{0.1cm}

    In this statement there is a misunderstanding of the words ``unit'' and ``unity'' as far as my earlier clarifications are concerned. While the word ``unit'' corresponds to some standard quantity, the word ``unity'' corresponds to the number ``one''. Then, the above quote symbolically\footnote{I can neither write ``mathematically'' nor write ``physico-mathematically''. So, I wrote ``symbolically'' even after knowing that verbal expressions are also symbolic. Here, I can not find any other option to express.} implies ``$e=1$'', which is meaningless because the left hand side is a quantity and right hand side is a number. What would have been meaningful is a statement like: {\color{OliveGreen}``charge $e$ could be made equal to unit''}. Considering the chosen quantity of charge, say $C$, as unit, the statement can be expressed, physico-mathematically,  as $e=C \equiv \epsilon_e=1$. However, if this is the case, then  $e$ is no more very small compared to $C$ i.e. $\epsilon_e=1\not\Leftrightarrow\epsilon_e\ll 1$ \footnote{The symbol ``$\not\Leftrightarrow$'' means ``does not imply each other''.}. Hence, a necessary condition (mentioned in the previous item) for Maxwell's definition of field  gets invalidated by Maxwell's own statement. Therefore, ``$F=Re$'' can not be written. So, the concept of field is not definable in the way Maxwell did. 
   	 
\end{enumerate}

\subsection{{Comments on Jackson's definition of electric field}}
 
 Now, I consider the definition of electric field as has been given by Jackson on page no. 24 of his book \cite{jackson}. It can be quoted as follows:\vspace{0.1cm}
 
 \begin{singlespace}
  {\small ``{\it  ....electric field can be defined as the force per unit charge acting at a given point... One must be careful in its definition, however. It is not necessarily the force that one would observe by placing one unit of charge on a pith ball and placing it in position. The reason is that {\color{blue}one unit of charge may be so large} that its presence alters appreciably the field configuration of the array. Consequently one must use a {\color{redish}limiting process} whereby the ratio of the force on {\color{blue}the small test body} to the charge on it is measured for {\color{redish}smaller and smaller amounts of charge}. Experimentally, this ratio and the direction of the force will become constant as the amount of {\color{redish}test charge is made smaller and smaller}.}''}    
 \end{singlespace}
\vspace{0.1cm}

Like that of Maxwell's case, Jackson's explanations are also based on ill-reasoned statements. I discuss those as follows.
 \begin{enumerate}
\item  {\color{blue}``{\it one unit of charge may be so large}''}:\vspace{0.1cm}

This statement is incomplete and meaningless because no mention of any quantity of charge has been made with respect to which the chosen ``unit of charge may be so large''.

\item {\color{redish}``{\it limiting process}''}:\vspace{0.1cm}

This statement has not been justified by any (physico-)mathematical explanation like an infinite series (with quantities), which generally comes to my mind when someone talks about limiting process. I shall elaborate on this particular issue shortly hereafter.

\item {\color{blue}``{\it the small test body}''}:\vspace{0.1cm}

There has been no mention of what aspect of the test body is small and with respect to what (similar to Maxwell's case). I assume that Jackson has meant ``mass'' of the object. Then, considering the mass of the test object to be $m$ and the standard mass to be compared with be $M$, such a statement is expressed as 
$m< M~\equiv~m=\epsilon_mM\ni~\epsilon_m< 1$.

\item {\color{redish}``{\it smaller and smaller amounts of charge}'', ``{\it test charge is made smaller and smaller}''}:\vspace{0.1cm}

Such statements are incomplete and meaningless because there is no clear explanation of ``smaller'' with respect to which charge. Further, there has been no clarification of whether such ``smallness'' has a limit (which I have discussed earlier). Although Jackson has mentioned in a footnote about such limit due to the experimental detection of discreteness of charge, then the question appears that whether the field due to that smallest charge is at all definable. This is a legitimate question to ask because the theoretical calculation that supports such detection of smallest charge, does involve the definition of field due to such ``smallest'' charge. I have already discussed this issue earlier.
 

\end{enumerate}
From the above clarifications it is clear in a rudimentary way how Jackson's definition of electric field is also ill-reasoned in a similar manner like that of Maxwell. However, unlike Maxwell, Jackson's clarification contain an important aspect, namely, the limiting process, which I discuss next.

\subsection{{Physico-mathematical expressions for the ``Limiting Process'' }}
Let me focus on the particular portion of Jackson's statement where he has written about a limiting process for the definition of electric field viz. `` {\it Consequently one must use ......... 
	charge is made smaller and smaller}.''

Here, I discuss the corresponding physico-mathematical expressions of such statement. So, let me consider a fixed charge unit $C$, a fixed mass unit $M$ and a fixed acceleration unit $a_0$ such that these result in a fixed force unit $F_0:=Ma_0$. Now, the limiting process can be interpreted as follows. First of all, assuming ``small test body'' implies ``small mass'', I complete such statement as follows
\begin{eqnarray}
m=\epsilon_mM\quad\ni \epsilon_m < 1
\end{eqnarray}
where $m$ is the mass of the test charge.
Then, I write the force felt by the test charge as 
\begin{eqnarray}
	F=ma=(\epsilon_m M)(\epsilon_aa_0)=\epsilon_FF_0\quad\ni\epsilon_F:=\epsilon_m\epsilon_a,F_0:=Ma_0.
\end{eqnarray}
Therefore, the expression for field should look like the following:
\begin{eqnarray}
E:=\lim\limits_{\substack{q\ll  C}}\frac{F}{q}=\lim\limits_{\substack{\epsilon_q\ll 1}}\frac{\epsilon_F}{\epsilon_q}\frac{F_0}{C}=\epsilon_E E_0\quad\ni E_0:=\frac{F_0}{C}, \epsilon_E:= \lim\limits_{\substack{\epsilon_q\ll 1}}\frac{\epsilon_F}{\epsilon_q}.
\end{eqnarray}
Now, let me recast the expression for $\epsilon_E$ as follows
\begin{eqnarray}
	\frac{E}{E_0}=\epsilon_E=\lim\limits_{\substack{\epsilon_q\ll 1}}\frac{\epsilon_F}{\epsilon_q}=\lim\limits_{\substack{\epsilon_q\ll 1}}\frac{\epsilon_m\epsilon_a}{\epsilon_q}\quad\ni\epsilon_m<1.
\end{eqnarray}
For some fixed $\epsilon_m<1$, if I choose $\epsilon_q$ smaller and smaller compared to unity, then for $\epsilon_E$ to tend to a constant value, say $\phi_0$, $\epsilon_a$ must have the general form as follows:
\begin{eqnarray}
\epsilon_a=\epsilon_q\epsilon_m^{-1}\(\phi_0+\sum_{n\geq 1}b_n[\epsilon_m]~\epsilon_q^n\),
\end{eqnarray}
 where $b_n$-s are some numbers which are essentially fixed by the choice of $\epsilon_m$. Then I can choose $\epsilon_q$ to be extremely small compared to unity, but not zero, such that I can write the following approximation
 \begin{eqnarray}
\lim\limits_{\substack{\epsilon_q\ll 1}} \epsilon_m^{-1}\(\phi_0+\sum_{n\geq 1}b_n[\epsilon_m]~\epsilon_q^n\)\simeq \frac{\phi_0}{\epsilon_m}
 \end{eqnarray} 
 to yield a satisfactory result that is useful for practical purpose\footnote{This is similar to what is done when the value of an irrational number, say $\sqrt 2$, is used for practical calculations and approximated according to requirement. Otherwise, it is a useless unknowable number that has an infinite continued fraction representation or a decimal representation with an infinite string of integers with a non-repeating pattern.}. What this last step implies is that {\it whatever quantity of charge I may choose, there always exists a smaller charge, in theory.} Such a statement has to be postulated as the premise to give a meaningful definition of electric field  which, unfortunately, has neither been discussed by Jackson, nor it is available in any existent literature (as far as my knowledge is concerned).

\subsection{Necessary premise for the definition of electric field { and the resolution of the contradiction:  undecidable charges as the middle way}}\label{premise}
Although I have briefly mentioned the necessary assumption to have a  definition of electric field, let me now declare it as a postulate\footnote{As I have already mentioned, since the physicist does not provide a definition of quantity like (\ref{chargedef}) and the subsequent clarifications regarding comparison of quantities (as in subsection (\ref{secconflict})), this postulate becomes necessary.}, followed by some crucial comments. \vspace{0.1cm}

{\bf Postulate:} {\it Considering a quantity of charge, written in terms of symbols to formulate a theory of physics, for any charge denoted by $q$, there always exists some charge $q'$ such that $q'<q$, or equivalently, $q'=\epsilon q~\ni 0<\epsilon <1$.}\vspace{0.1cm}

By this postulate, the test charge for the definition of electric field can be chosen arbitrarily small compared to the chosen unit charge so that the limiting process is possible. When the experiment yields that the smallest experimentally detectable charge among all other detectable ones, then by the above postulate the theoretical definition of the electric field is still possible. It only requires an added clarification that there exist charges ($Q_u$-s), by postulate, such that for any $Q_u$ the following condition holds: $Q_u<e$, or equivalently,  $Q_u=\epsilon_u e~\ni 0<\epsilon_u<1$. This justifies, on logical grounds, the use of electric field in the theory that explained the results of the oil drop experiment in ref.\cite{millikan} {\color{black} and therefore, resolves the contradiction that I pointed out in subsection (\ref{oildrop}).} I call such $Q_u$-s as {\it undecidable}\footnote{The word `undecidable' is borrowed from Goedel \cite{goedelincom}. However, my reasoning does not rely on the formal language of mathematical logic and rather may be considered as ``sane process of reasoning''\cite{logicindia} that seems useful, practical and serves the purpose.} charges. Since $e$ represents the charge such that any charge smaller than $e$ can not be experimentally detectable and hence, the existence of such charges ($Q_u$-s) can neither be proven true nor be proven false i.e. existence of $Q_u$-s is experimentally undecidable.

Now, there remains the question of theoretical undecidability i.e. if the postulate is regarded as ``hypothesis'', then whether this ``hypothesis'' can be proven to be true or false within the theory. This question can be answered as follows. I have  ``hypothesized'' $Q_u$-s to define electric field through a valid physico-mathematical limiting process without having any logical fallacies (circular traps or useless contradictions) and this provides a justification for the use of electric field in the theory associated with the oil drop experiment which confirmed observation of discrete charge. Thus, the entire construction is built upon the ``hypothesis'' of $Q_u$-s. Now, if one wonders whether there is a theoretical proof or disproof of this ``hypothesis'', the answer is obviously negative. This is because the ``hypothesis'' is the only reason that has let me write down the theory itself by avoiding any logical fallacy. Any trial to prove or disprove the ``hypothesis'' theoretically is like using the premise to prove or disprove itself. A successful proof of the ``hypothesis'' leads to a useless tautology: 
\begin{eqnarray}
	\text{{\it $Q_u$ exists theoretically, if the existence of $Q_u$ is hypothesized to construct the theory}.}\nonumber
\end{eqnarray} 
And, a successful disproof of the same leads to a useless contradiction:
\begin{eqnarray}
\text{{\it $Q_u$ can not exist theoretically,  if $Q_u$ is hypothesized to construct the theory.}}\nonumber
\end{eqnarray} 
Therefore, the ``hypothesis'' is theoretically neither provable nor unprovable i.e. undecidable. This is why I have used the word ``postulate'' rather than the word ``hypothesis''. {\color{black} Finally, to draw the connection with what I have discussed in subsection(\ref{nonbody}), I may state that the undecidable charges provide the middle way -- the founding premise -- based on which the truth of the {\it experimentally detectable} smallest charge, among all experimentally detectable charges, is verified based on the theory that requires arbitrarily small {\it theoretically defined} charge compared to any other charge. It is clearly manifest that removing the adjectives ``experimentally detectable'' and ``theoretically defined'' brings back the contradiction again that I discussed in subsection(\ref{oildrop}).}

\subsection{Maxwell's relational statements{: the essence of dependent origination and emptiness}}\label{maxwellrelation}
  There are two sentences in Maxwell's explanations regarding his concept of electric field on page no. 45 of ref.\cite{maxwell} which need separate attention. This is because those  two sentences could have led physics along a different path of evolution if the same were carefully written in terms of physico-mathematical expressions {\color{black} and the essence of dependent origination and emptiness  would have been profoundly manifest}. I explain my view as follows by quoting the respective sentences one by one.\vspace{0.1cm}
 
Sentence 1: {\small ``{\it The Electric Field is the portion of space in the neighbourhood of electrified bodies, considered with reference to electric phenomena.}''}\vspace{0.1cm}

This sentence is suggestive of the fact that Maxwell considered the notion of space with reference to charge. This is because ``charge'' is the most elementary expression with which ``electric phenomena'' are explained, as Maxwell did. On the other hand, the thought of ``space'', ``neighbourhood'', etc. give rise to, are or associated with, the notion of ``distance'' which, however, needs to be explained in terms of physico-mathematical expressions. If the thought of ``distance'' is conceived with reference to ``charge'', then certainly those two should be comparable. And, then both should be of the same physical dimension. Given that ``distance'' is just a thought of an extension and represented as having length dimension, ``charge'' should also be considered to have length dimension i.e. a {\it type} of extension \footnote{In ref.\cite{essay}, it has been shown how to write down a non-singular theory of gravity by expressing the existence of an object in terms of an intrinsic extension instead of ``mass''. Maxwell's statements are suggestive of a similar operation with charge as well. {\color{black} However, this would have a significant effect on how the units of electricity are fixed through practical methods based on chemical processes like  electro-deposition of metals (or volumes of gas produced). Since this is the most substantial empirical basis for the existence of units of electricity, considering the change in the language through which the expressions are made, along with the associated physico-mathematical analysis, there needs to be a thorough revision of the ideas. Notably, from the logico-linguistic point of view, this brings in the question regarding the categorical distinction, namely, ``chemical'' and ``physical''. Nevertheless, such analysis is beyond the scope of the current text.}}. In symbolic terms, this implies
\begin{eqnarray}
 [Q]\equiv [L].
\end{eqnarray}
 Otherwise, the two concepts can not be related or compared to each other so that Maxwell could think of ``distance'' with ``reference'' to ``charge''. If such formulation of a theory for electric phenomena is penned then the distance $d$ can be compared to both the source charge $q_s$ and test charge $q_t$. Therefore, if one can show that the Coulomb's law follows only in the condition 
 \begin{eqnarray}
  d\gg q_s,q_t,
 \end{eqnarray}
 then it actually provides a justification for {\color{black} the modern usage of the word ``macroscopic'' in order to identify basic electrostatics, }which is otherwise unexplained and ill-reasoned\footnote{\color{black}For example, Jackson writes, in a footnote, on page no. 25 in ref.\cite{jackson} while explaining the limiting process for the test charge --``{\it This is an example of a mathematical idealization in  {\color{redish}macroscopic} electrostatics.}''. I find the use of the word ``macroscopic'' highly objectionable due to the following reasons. As far as I understand, the word ``macroscopic'' is associated with length units. When observed phenomena at the conventional units like ``metre'', ``centimetre'', etc. are studied (that is visible to the naked eye), the phenomena are called macroscopic. It is not at all understandable where the involvement of length units come from, because what have been involved in the discussion regarding the limiting process are force, charge and mass. One may immediately refer to the literature of {\it quantum} electrodynamics and the role of length (resolution) units in the process of renormalization. Then, the question arises whether one needs to study {\it quantum} electrodynamics and renormalization in order to understand the definition of {\it classical} electric field. Also, Millikan and Fletcher did not need to consider any renormalization scheme or {\it quantum} electrodynamics to write down the theory which founded the basis of their experiment. Therefore, I find such reasoning to be absurd and not worthy of further discussion just for the sake of destructive criticism. And besides that, as far as the credibility of {\it quantum} electrodynamics is concerned, the foundations are still very embarrassingly questionable  \cite{rottenqed}.}. This is because here the word ``macroscopic'' can now be explained by the behaviour of the theory when ``the distance is large  compared to both the charges''. I hope to report such expositions elsewhere. {\color{black} However, what is manifest from Maxwell's statement is that the notions of ``distance'' and ``charge'' originate in a mutually dependent fashion i.e. neither ``distance'' nor ``charge'' can be realized independent of each other. Hence, each of these categories are empty of truth in itself and only true in relation to each other, while both arise in relation to the observer.}\vspace{0.1cm}


Sentence 2:  {\small ``{\it It may be occupied by air or other bodies, or it may be a so-called vacuum, from which we have withdrawn every substance which we can act upon with the means at our disposal.}''}\vspace{0.1cm}

This sentence suggests that Maxwell conceived of the notion of ``vacuum'', or synonymously ``emptiness'', in {\it relation} to the observer's (``our'') ability rather than an {\it absolute} concept. Since the role of observer is manifested through the choice of units for measurement, then certainly the physico-mathematical expressions of such thought  should involve the ability to choose the units. Also, the question arises -- ``vacuous or empty of what? mass or charge or both?'' The word ``every substance'' is suggestive of the fact that Maxwell meant both ``charge'' and ``mass''. Further, the word ``act upon'' is suggestive of the involvement of ``force'' in the conception of ``vacuum''.
Therefore, the concept of such ``relative vacuum or emptiness'' should be expressed in terms of relational statements about the respective quantities like the following $m\ll M, q\ll C, F\ll F_0$ such that these correspond to extremely small numbers compared to unity like $\epsilon_m,\epsilon_q,\epsilon_F\ll 1$ for the respective cases. Here, $M,C,F_0$ represent ``the means at our (observer's) disposal'', $m,q$ represent ``every substance'' and $F$ represents the action that is beyond realization of the observer who acts to create the vacuum by the ``means (of action) at disposal'' ($F_0$).

{\color{black} Unfortunately, no effort, such as the one explained above, was made by Maxwell to convert his verbal expressions to physico-mathematical ones. Needless to say, an implementation of the concept of ``relational vacuum''  would have given a differently structured equations rather than those known as Maxwell's equations. Absence of such an analysis resulted in an evolution towards modern physics which inherits only the concept of ``absolute vacuum'' in the context of field theory albeit in a more sophisticated fashion. Such ``absoluteness'' is robustly established as a geometric invariant under space-time translation and Lorentz transformations e.g. see ref.\cite{coleman}. }

  \section{Outlook}\label{summary}
{\color{black} I conclude with the provision of an outlook through a summary of what I have discussed in this work from which the reader may be able to understand in what sense this work can lead to some hitherto unattended pastures of scientific thoughts and why the same is associated with ample risk of misinterpretations. First and foremost, the risk can be understood immediately from the Indian philosopher's point of view, as Dasgupta wrote in the {\it Preface} of ref.\cite{surendranath}:\vspace{0.1cm} 
	
{\small ``{\it Much harm has already been done by the circulation of opinions that the culture and philosophy of India was \underline{dreamy and abstract}.......It is unfortunate that the task of re-interpretation and re-valuation of Indian thought has not yet been undertaken on a comprehensive scale.}''}\vspace{0.1cm}
	
	 I believe, however, the motivation to take on such a task can be ignited only through a demonstration of how such thought process can directly affect science and especially physics. This work is an attempt to showcase such a demonstration, with an appeal for the dissociation of the categorical disciplines of logic and physics and on the large, a fruitful merger of Eastern philosophy and Western science. The subtlety of this work, of course, is manifested in the discussion regarding the ``I'' to begin with and then the passage through different layers of self-inquiry can make it a bit awkward to the reader who is not habituated with the same. I have really taken the risk of being criticized as ``dreamy and abstract'' by ignoring Radhakrishnan's warning, in the preface of ref.\cite{radhakrishnan}, against the modern mind's judgment of Indian philosophy as  ``{\it two or three `silly' notions about `maya', or the delusiveness of the world, `karma', or belief in fate, and `tyaga', or the ascetic desire to be rid of the flesh.}'' Nevertheless, the truth of the closed line as the middle way, that provides the ground for the judgment of what is area and what is not, is a directly demonstrated truth that can be immediately realized through self-inquiry by even the most naive reader just by asking himself, ``What am I seeing?''. Yet the essence of the example lies beyond formal logic that is founded on the law of excluded middle. This should not be misinterpreted as the ``violation'' of the law of excluded middle, but it is a demonstration of the fact that the middle, that is only excluded by choice, is the premise that lets the observer realize, through the modes of observation, the distinctive categories and the realization of the premise itself -- the middle way -- occurs through self-inquiry (see ref.\cite{companion} for a comparison with Frege's metaphorical demonstration). Buddhism in particular, and Indian philosophy in general, relies on such direct reasoning that is closely attached to experience. Therefore, the associated thought process lies in the middle of being completely logical (that excludes the middle way) and being completely operational (that relies often on untruthful expressions of experience), which can be aptly considered as ``sane process of reasoning''\cite{logicindia}. Avoidance of the subtlety, associated with such sane process of reasoning attached to experience, has led the physicist of the modern days to the conception of  ``dreamy and abstract'' desires of writing down ``the final theory''\cite{weinberg} or ``the theory of everything''\cite{hawking} or of constructing ``the road to reality'' that provides ``a complete guide to the laws of the universe''\cite{roadtoreality}. Such ``dreamy and abstract'' desires of the physicist are not only devoid of self-inquiry but also founded on unexplained act of treating physical dimension as number and logico-linguistically fallacious statement of ``zero quantity''. 
	 
	     It is through self-inquiry, where one investigates the truthfulness of the expressions of his own perceptions or observations, the dependent origination of quantity, number and physical dimension becomes apparent. Also, the essence of the middle way can be now understood in the use of physical dimensions in physics e.g. physical dimension is not used as number while defining quantity but it needs to be used as number for the explanation of addition of two quantities. The relational existence of quantity provides the path to truthful expressions which are devoid of the logico-linguistic fallacy of ``zero quantity''. Now,} only relational statements are considered meaningful, such as, ``a quantity smaller than another quantity $(q_1<q_2\equiv q_1=\epsilon_{12} q_2\ni \epsilon_{12}<1)$'', ``a quantity is extremely or infinitesimally smaller than another quantity $(q_1\lll q_2\equiv q_1=\epsilon_{12} q_2\ni \epsilon_{12}\lll 1)$'', etc., {\color{black}where the verbal statements have a truthful correspondence to the physico-mathematical statements.} Although such relational statements  appear to be the most basic lessons one gets while learning about measurements and units (chosen standard quantities) in physics, the empirical essence of such statements are quickly forgotten while writing down the most basic theories, {\color{black} as the physicist does not make self-inquiry}. For example, the physicist's definition of electric field depends on the statement ``$q_1\to 0$'' where $q_1$ is a test charge. Two kinds of verbal explanations can be found in the literature for the statement ``$q_1\to 0$'' and none of those explanations verbally justifies ``zero quantity'' in relation to the physicist's definition of electric field. Rather both of those explanations depend on verbal expressions that manifest additional concepts than only ``zero quantity'': (i) One of those explanations depends upon the use of the concept of electric field due to the test charge rather than the test charge itself. Therefore, in order to define field (due to the source charge), the concept of field (due to the test charge) is considered as predefined. Hence, such definition by the physicist becomes circular in reasoning. (ii) The other explanation depends upon the comparison between   the source charge $(q_2)$ and test charge $(q_1)$. Such explanation is clearly misleading in the first place because ``$q_1\to 0$'' does not contain any explicit reference to $q_2$. Replacement of ``$q_1\to 0$'' with the statement ``$q_1\ll q_2 \equiv q_1=\epsilon_{12} q_2\ni\epsilon_{12}\ll 1 $'' leaves the physicist's definition of electric field incomplete {\color{black}because it ends up in an unresolved contradiction concerning the Millikan-Fletcher oil drop experiment.} The explanation of ``$q_1\to 0$'' in the existing literature is an admixture of both (i) and (ii). It is especially interesting to take note of the mismatch between verbal expressions and physico-mathematical statements of Maxwell in ref.\cite{maxwell} and of Jackson in ref.\cite{jackson}. It is even more interesting to observe how a truthful conversion of both Maxwell's and Jackson's verbal statements to the corresponding physico-mathematical ones could have led to a different kind of analysis based on relational statements like $q_1\lll q_2$. {\color{black} The essence of self-inquiry and relational existence becomes apparent through such analysis. 
     
         Especially Jackson's limiting process regarding the test charge indicates the necessity of a postulate regarding the existence of charges which nevertheless follows from the definition of quantity in terms of number and physical dimension which is devoid of the concept of ``zero quantity''. Also, while such a postulate renders the electric field definable, it also resolves the logical dilemma that arises otherwise in the context of the oil drop experiment. The concept of undecidable charges ($Q_u$-s) play the pivotal role in the process. That is, if $e$ is the smallest detectable charge then $Q_u<e$ renders the electric field due to $e$ as definable and hence, definable for any charge. So, the existence of $Q_u$ is neither provable nor disprovable in theory as the theory itself is constructed on the basis of its theoretical definition. And, $Q_u$ is experimentally neither verifiable nor falsifiable because $e$ is the smallest detectable charge among all other detectable charges and it is the theory that makes it possible to reach a conclusion about $e$. The undecidable charges act as the middle way that lets the judgments regarding the theory and the experiment possible. Further, a careful investigation of Maxwell's verbal statements point towards dependent origination of distance and charge and a relational concept of vacuum. Due to Maxwell's comparisons of charge and distance, a truthful conversion of such verbal statements to physico-mathematical statements requires an identification of charge dimension and length dimension. And, the relational concept vacuum is a direct manifestation of emptiness, which essentially is in conflict with the modern concept of absolute vacuum.}     
   
   {\color{black} In a nutshell, the logico-linguistic inquiry that I have intended to present here is much in the spirit of arithmetic reasoning adopted by the logician, but it retains the doubt that is excluded, by choice, by the logician owing to the exclusion of the middle and the lack of self-inquiry. Possibly the most profound implication of the whole analysis, in short, is the abolition of ``zero quantity''. This brings in questions regarding the foundations of calculus as applied to physics. Certainly while Indian philosophy, especially the notion of relational existence of Buddhism, founds the basis this discussion, a spontaneous curiosity may arise regarding the literature of mathematical science associated with such cultures (e.g. see \cite{yuktibhasa,aryabhatiyaclark,calculusindia} and many others not to be cited here), which nevertheless was devoid of the notion of physical dimensions and hence, physics as known today. The root of calculus was devoid of the concept of {\it limit} and rather was founded on infinite series expansions (although restricted to trigonometric functions only) which were truncated for practical use and never written as an exact result unlike what is done in modern calculus.  This is very similar to the expression of the electric field, due to the application of the ``limiting process'' of Jackson, which involves $\epsilon$-s i.e. very small numbers. Keeping the connection with Indian philosophy aside, such analysis also appears to be similar to non-standard analysis in structure. Such a subject, established by Robinson\cite{robinsonmeta,robinson} on the basis of  mathematical logic, however, provides no clarification regarding the concepts of physical dimension, number, quantity, unit and unity.

   	Nonetheless, the question remains whether it is possible to restructure physics from the scratch based on such a transformed philosophical base and, most importantly, what new insights one can gain from such a paradigm shift. With an intention of answering such questions, I intend to continue with further investigations regarding the foundations of calculus, a glimpse of which can be found in ref.\cite{companion} and more elaboration in refs.\cite{comment,essay}. To me, at least, a more reasonable physico-mathematical science with new philosophical underpinnings  seems upcoming.} 
     
{\it Acknowledgment:} {\footnotesize The author thanks G. Sardar for helpful comments on an initial draft of this paper and the students of B. Stat. 2nd year, 2019-2020, of Indian Statistical Institute, Kolkata, especially T. Mondal and T. Shri, for helping with their scholarly inquisitive attitude. This work is supported by  the Department of Science and Technology of India through the  INSPIRE Faculty Fellowship, Grant no.- IFA18-PH208.}

   \begin{appendices}
    
    \section{Remarks on Russell's view of number, quantity, etc.}\label{appb}
    In this regard I find it relevant to consider the view of Russell ref.\cite{russellnum}:\vspace{0.1cm}
    
    {\small {\it ``Since a unit must be defined by some quality, number will thus have no reference to a unit, or rather its unit is the abstract object of any act of attention, of whatever kind this may be. Such an operation can only give rise to the natural numbers, the series of positive integers.''}}\vspace{0.1cm} 
    
    There is  partial agreement and partial disagreement of Russell's view with that of mine. The agreement is on two occasions. Firstly, I agree with Russell on the view that number itself is devoid of any reference to a unit and this is what I have tried to convey while explaining the difference between {\it unit} and {\it unity}. Secondly, I agree that  measurement gives rise to positive real integers.  The disagreement is for two reasons. Firstly, Russell related ``unit'' to ``quality'' while I considered ``quantity'' and secondly, Russell considered exact measurement while I have considered inexact measurement. The second disagreement is same as that I have with Einstein. 
     
     Given that I have mentioned Russell's work in course of my discussion, there is a responsibility on my behalf to make a clear statement regarding the distinction between ``quality'' and 	``quantity'', which I aim to do in some future work if I get the opportunity. Nonetheless, as far as  physical dimensions are concerned, I do not find any discussion similar to that of mine anywhere in the literature (as far as my knowledge is concerned). This leads me to think whether the ``quality-quantity'' distinction can also be addressed through a middle-way. While I write this, I am completely aware of Russell's attempt to resolve such contradiction in ref.\cite{russellnum}. I am also aware of Russell's use of the word ``infinitesimal'' in association with ``quantity'' in ref.\cite{russellnum} and whether ``infinitesimal quantity'' is a meaningful phrase at all is being questioned  in this present discussion.
     
     \section{Comments for the logician}\label{logicians}
     	The current piece of work is nowhere near the form that a logician may like because I have not used the specific jargon of mathematical logic to discuss the issues at hand. Rather my intention is to present a discussion by adopting a middle way, which is neither that of the logician nor that of the physicist. It is important to note that the passage from Tarski's work that I have quoted in subsection (\ref{logicdef}), is only intended to reflect the attitude of the logician in general i.e. the logician wishes to achieve completeness in logic such that it ``admits of no doubt''. It is such attitude that is under discussion. Interestingly, if the basic expressions of the physicist -- physical dimensions -- are judged with the logician's attitude, then the very foundations of physics seems illogical. 
     	
     	Nevertheless, the logician may wonder whether it is necessary, for the present purpose, to delve deeper into the specific jargon and notations of mathematical logic that has been used by Tarski to explicate the meaning of ``definition'' (e.g. in Chapter 10 of ref.\cite{tarskiundef}). The answer depends on what one wants to do. I intend to keep the discussion devoid of unnecessary abstraction which often glosses over the main issue at hand and it is simply because the questions regarding physics, that I discuss here, are so elementary that the conversions of the verbal expressions to physico-mathematical expressions can be done directly and with an immediate effect on physics. The logician, who seeks a packaging of this work within the folds of the specific jargon and the abstractions of mathematical logic just for the sake of sophistication, must clarify a practical motive of doing so and also he must justify what he gains over this current piece of work in adopting such a method e.g. ref.\cite{comment,essay,companion} suggest how simple logico-linguistic inquiry regarding the definition of derivative can change the way calculus and physics are done today.

     	Now, as far as admitting doubt (or incompleteness) in the meaning of  ``definition'' and my comparison with Tarski is concerned, I may point out the following. On page no. 297 of ref.\cite{tarskiundef}, Tarski clearly states the domain of his work through the following statement:\vspace{0.1cm}
     	
     	{\small ``{\it We shall be concerned here only with those deductive theories which are based upon a sufficiently developed system of  mathematical logic. Problems concerning concepts of logic itself will not be considered. ....... I shall consider exclusively the scientifically constructed languages known at the present day, i.e. the formalized languages of the deductive sciences. }''.}\vspace{0.1cm}

     	 Physics, neither is completely deductive (as it requires induction, empirical inputs, even intuitive guesses sometimes, etc.), nor it has been written in a formalized language. In fact this has been pointed out by Poincare and also to some extent, albeit indirectly, by Dedekind in the context of calculus (see footnote no. 8). Therefore, I do not see a necessity of using the specific jargon and the abstract notations of mathematical logic in the present context. For further discussion and comments the reader may consult ref.\cite{companion}.

\end{appendices}

{\small{\it Acknowledgments:} The author thanks G. Sardar for discussions and being supportive. This work is supported by the Department of Science and Technology of India through the INSPIRE Faculty Fellowship, Grant no.- IFA18-PH208.}


\begin{thebibliography}{777}

\bibitem{poincareq1} Quoting Poincare from page no.6 of ref.\cite{poincare}: ``{\it We must seek mathematical thought where it has remained pure - i.e., in Arithmetic.}''

\bibitem{poincare} H. Poincare, {\it Science and Hypothesis}, The Walter Scott Publishing Co. Ltd., New York (1905); \href{https://www.gutenberg.org/files/37157/37157-pdf.pdf}{Project Gutenberg link.}


\bibitem{fregeq1} Quoting Frege from page 21 of ref.{\cite{frege1}}: {\it `` The basis of arithmetic lies deeper, it seems,
	than that of any of the empirical sciences, and even than that of geometry. The truths of arithmetic govern all that is numerable. This is the widest domain of all; for to it belongs not only the actual, not only the intuitable, but everything thinkable.''}

\bibitem{frege1} G. Frege, {\it The Foundations of Arithmetic}, Northwestern University Press (1980), translated into English by J. L. Austin. 


\bibitem{weyl} It is interesting to note Weyl's acknowledgement of the fact that {\it ``the concept of number appears as logically prior to the concepts of geometry''} in course of addressing the nature of evolution of mathematics over the history of science; see page (viii) of ref. \cite{weyl1}. 

\bibitem{weyl1} H. Weyl, {\it The theory of groups and quantum mechanics}, Kessinger Publishing, LLC (2008), translated into English from the second German edition by  H. P. Robertson. 

\bibitem{ifrah1} Quoting  Ifrah from page (xxii) of ref.\cite{ifrah2}: {\small {\it `` For Plato, numbers
	were ``the highest degree of knowledge'' and constituted the essence of  outer and inner harmony. The same idea was taken up in the Middle  Ages by Nicholas Cusanus, for whom ``numbers are the best means of approaching divine truths''. These views all go back to Pythagoras, for 	whom ``numbers alone allow us to grasp the true nature of the universe''. In truth, though, it is not numbers that govern the universe. Rather, there are physical properties in the world which can be expressed in
	abstract terms through numbers. Numbers do not come from things themselves, but from the mind that studies things.''}}

\bibitem{ifrah2}G. Ifrah, {\it  The Universal History of Numbers: From Prehistory to the Invention of the Computer}, Wiley (2000), translated into English by  D. Bellos, E. F. Harding, S. Wood and I. Monk.


\bibitem{lilavati} Bhaskaracarya, {\it Lilavati of Bhaskaracarya: A Treatise of Mathematics of Vedic Tradition}, Motilal Banarsidass Publishers Pvt. Ltd. (2001); translated by K. S. Patwardhan, S. A. Nampally, S. L. Singh.


\bibitem{peano}G. Peano, {\it The principles of arithmetic, presented by a new method } (1889) on page no. 101 in  {\it Selected works of Giuseppe Peano}, translated by H. C. Kennedy, University of Toronto Press (1973).


\bibitem{principia1} I. Newton, {\it  Principia, Volume I - The Motion of Bodies}, University of California Press (1966). [English translation by A. Motte, edited by F. Cajori]

\bibitem{principia2} I. Newton, {\it  Principia, Volume II - The System of the World}, University of California Press (1966). [English translation by A. Motte, edited by F. Cajori]


\bibitem{maxwell}J. C. Maxwell, {\it A Treatise on Electricity and Magnetism}. Volume 1, Clarendon Press (1873).

\bibitem{bridgman} P. W. Bridgman, {\it Dimensional Analysis}, Yale University Press (1963).

\bibitem{dimhis} See refs.\cite{dim1,dim2} and the references therein for a historical account of the literature concerning dimensional analysis in physics.

\bibitem{dim1} E. O. Macagno, {\it Historico-critical review of dimensional analysis}, Journal of the Franklin Institute, 
Vol. 292, Issue 6, pp. 391-402 (1971), \href{https://doi.org/10.1016/0016-0032(71)90160-8}{online link.}

\bibitem{dim2}R. A. Martins, {\it The origin of dimensional analysis}, Volume 311, Issue 5, May 1981, Pages 331-337, \href{https://doi.org/10.1016/0016-0032(81)90475-0}{online link.}


\bibitem{nist}NIST Special Publication 330, The International System of Units (SI),  \href{https://nvlpubs.nist.gov/nistpubs/SpecialPublications/NIST.SP.330-2019.pdf}{online link.}


\bibitem{einphil2}  A. Einstein, {\it  Relativity: The Special and the General Theory}, Forgotten Books (2010).


\bibitem{russellnum}B. Russell, {\it On the Relations of Number and Quantity},
{\it Mind}, New Series, Vol. 6, No. 23, pp. 326-341 (1897),  Oxford University Press on behalf of the Mind Association, \href{http://www.jstor.org/stable/2247724}{online link}.


\bibitem{massgap}A. Jaffe, E. Witten, {\it Quantum Yang-Mills theory}, \href{https://www.claymath.org/sites/default/files/yangmills.pdf}{online link to the official statement of the problem}.


\bibitem{navsto}C. Fefferman,  {\it Existence and smoothness of Navier-Stokes equation}, \href{https://www.claymath.org/sites/default/files/navierstokes.pdf}{online link to the official statement of the problem.}

\bibitem{michell}J. Michell, {\it History and philosophy of measurement: a realist view},  \href{https://www.imeko.org/publications/tc7-2004/IMEKO-TC7-2004-128.pdf}{online link}.

\bibitem{stanmea}{\it Measurement in Science}, Stanford Encyclopedia of Philosophy, \href{https://plato.stanford.edu/entries/measurement-science/}{online link}.

\bibitem{wolff}J. E. Wolff, {\it The metaphysics of quantities}, Oxford University Press (2020).

\bibitem{dedekindnum} R. Dedekind, {\it Essays on the Theory of Numbers: (I) Continuity and Irrational Numbers (II) The Nature and Meaning of Numbers}, Dover Publications (1963).

\bibitem{cauchycal} D. M. Cates, {\it  Cauchy's Calcul Infinitesimal}, Springer International Publishing (2019).

\bibitem{essay} A. Majhi, {\it Contradictions, mathematical science and incompleteness}, \href{https://www.researchgate.net/publication/340499727_Contradictions_mathematical_science_and_incompleteness}{online link}.

\bibitem{comment} A. Majhi, {\it Comments on the applications of calculus in physics}, \href{https://www.researchgate.net/publication/340717863_Comments_on_application_of_calculus_in_physics/stats}{online link}.


\bibitem{heijen}  J. Heijenoort, {\it From Frege to Goedel - A Source Book in Mathematical Logic, 1879-1931}, Harvard University Press (2002).


\bibitem{bragg} W. B. Ewald, {\it From Kant to Hilbert: A source book in the foundations of mathematics, Vol. 1 \& 2}, Oxford University Press, USA (2007).

\bibitem{gilliesmath} D. Gillies, {\it  Frege, Dedekind and Peano on the Foudations of Arithmetic}, Van Gorcum (1982).

\bibitem{fraenkel} A. Fraenkel, Yehoshua Bar-Hillel, Azriel Levy, {\it Foundations of Set Theory - Studies in Logic and The Foundations of Mathematics, Volume 67}, Elsevier (1973).

\bibitem{cantor1} E. Huntington,  G. Cantor, {\it The continuum, and other types of serial order - with an introduction to Cantor's transfinite numbers}, Dover Publications (2003).
	
\bibitem{cantor2} G. Cantor, {\it  Contributions to the founding of the theory of transfinite numbers}, Nabu Press (2010).


\bibitem{pm1} A. Whitehead, B. Russell,  {\it Principia Mathematica},  Vol. 1, Cambridge University Press (1927).

\bibitem{goedelincom}K. Goedel, {\it On formally undecidable propositions of Principia Mathematica and related systems I} (1931) in  {\it Kurt Goedel, Collected Works, Volume 1, Publications 1929-1936}, edited by  S. Feferman, J. W. Dawson Jr., S. C. Kleene, G. H. Moore, R. M. Solovay, J. Heijenoort, Oxford University Press, New York; Clarendon Press, Oxford (1986).


\bibitem{aristotlepa} Aristotle, {\it The Complete Works of Aristotle (The Revised Oxford Translation)}, Princeton University Press (1984).


\bibitem{logichandbook8} D. M. Gabbay, J. Woods (eds.), {\it Handbook of the History of Logic, Volume 08: The Many Valued and Nonmonotonic Turn in Logic}, Elsevier (2007).

\bibitem{multifuzzy} M. Bergmann, {\it An Introduction to Many-Valued and Fuzzy Logic: Semantics, Algebras, and Derivation Systems}, Cambridge University Press (2008).


\bibitem{fregeoneunity} Frege wrote on  page no. 58 of ref.\cite{frege1}: {\it ``A distinction must be drawn between one and unit.''}. 

\bibitem{fuzzy1} R. R. Yager, L. A. Zadeh (eds.), {\it An Introduction to Fuzzy Logic Applications in Intelligent Systems}, Kluwer Academic Publishers (1992).



\bibitem{fuzzy2} T. J. Ross, {\it Fuzzy logic with engineering applications}, John Wiley \& Sons, Ltd., (2004).


\bibitem{logichandbook3} D. M. Gabbay, J. Woods (eds.), {\it Handbook of the History of Logic, Volume 03, The Rise of Modern Logic: From Leibniz to Frege}, Elsevier (2004).

\bibitem{nagel} E. Nagel, {\it The Structure of Science: Problems in the Logic of Scientific Explanation}, Harcourt, Brace \& World, INC. (1961).

\bibitem{popper} K. Popper, {\it The logic of scientific discovery}, Routledge (2002).

\bibitem{wittgenstein} L. Wittgenstein, {\it Tractatus Logico-Philosophicus}, Routledge (2001).

\bibitem{duhem} P. Duhem, {\it The Aim and Structure of Physical Theory}, Princeton University Press (1954). 

\bibitem{turing} A. Turing, {\it Systems of Logic Based on Ordinals}, Proceedings of the London Mathematical Society (1939).

\bibitem{wallace} B. A. Wallace, {\it Buddhism and Science}, Columbia University Press (2003).

\bibitem{tarskiundef} A. Tarski, 
{\it Logics, Semantics, Metamathematics - Papers from 1923 to 1938}, translated by J. H. Woodger, Second Edition, Hackett (1983).

\bibitem{tarskidef} A. Tarski, {\it Introduction to logic and to the methodology of the deductive sciences}, Oxford University Press (1994).

\bibitem{logicphysics} P. W. Bridgman, {\it The Logic of Modern Physics}, The Macmillan Company (1960).


\bibitem{boole}G. Boole, {\it An investigation of the laws of thought, on which are founded the mathematical theories of logic and probability}, Walter and Maberly (1854).


\bibitem{nayak} G. C. Nayak, {\it  Madhyamika Sunyata: A Reappraisal}, Indian Council of Philosophical Research (2001).

\bibitem{garfield} J. Garfield, {\it The Fundamental Wisdom of the Middle Way: Nagarjuna's Mulamadhyamakakarika}, Oxford University Press, USA (1995).

\bibitem{siderits}  M. Siderits, S. Katsura, {\it Nagarjuna's Middle way -- the Mulamadhyamakakarika (Classics of Indian Buddhism)}, Wisdom Publications (2013).

\bibitem{stre} F. Stcherbatsky, {\it  Buddhist Logic} (two volumes), Dover Publications (1962).

\bibitem{bodhi} Bhikkhu Bodhi, {\it The Great Discourse on Causation: The Mah$\bar{a}$nid$\bar{a}$na Sutta and Its Commentaries}, Buddhist Publication Society (1995).

\bibitem{logicindia} As Potter would write, ``{\it Indian logic is never conceived as `formal' in the Western sense, but as an account of sane processes of reasoning it has few equals in the West for attention to detail.}'', on page no. 2 of ref.\cite{potter2}.

\bibitem{potter2} K. Potter (editor) - {\it Encyclopedia of Indian Philosophies, Volume II -- Indian Metaphysics and Epistemology}, {\it The Tradition of Nyaya-Vaisesika up to Gangesa}, Motilal Banarsidass (1995).


\bibitem{surendranath} S. Dasgupta, {\it  A History of Indian Philosophy, Vol. I}, Cambridge University Press (1922).

\bibitem{radhakrishnan} S. Radhakrishnan, {\it Indian Philosophy}, Vol. 1, George Allen \& Unwin (1948).

\bibitem{sinha} J. Sinha, {\it Indian Philosophy}, Vol. 1, 2 \& 3, Motilal Banarsidass (2016).

\bibitem{descartes} R. Descartes, D. E. Smith, M. L. Latham, {\it The Geometry of Rene Descartes}, Dover Publications, Inc. (1954).

\bibitem{hawking} S. W. Hawking, {\it The theory of everything: The Origin and Fate of The Universe}, Phoenix Books (2006).

\bibitem{weinberg} S. Weinberg, {\it Dreams of a final theory: The Scientist's Search for the Ultimate Laws of Nature}, Vintage Books (1994).

\bibitem{roadtoreality} R. Penrose, {\it The Road to Reality: A Complete Guide to the Laws of the Universe}, Jonathan Cape (2004).


\bibitem{landau}L. Landau,  E. Lifshitz, {\it The Classical Theory of Fields - Course of Theoretical Physics  Volume 2}, Butterworth Heinemann (1994).

\bibitem{coleman}S. Coleman, {\it Quantum Field Theory: Lectures of Sidney Coleman}, World Scientific (2019).

\bibitem{analyst} G. Berkeley, {\it The Analyst} (1734) [edited by D. R. Wilkins],  \href{https://www.maths.tcd.ie/pub/HistMath/People/Berkeley/Analyst/Analyst.html}{online link}.

\bibitem{measure} T. M. Apostol, M. A. Mnatsakanian, {\it New horizons in geometry}, The Mathematical Association of America (2013).


 
 \bibitem{fom1} D. H. Krantz, R. D. Luce, P. Suppes, A. Tvesky, {\it Foundations of measurement: Volume I - Additive and polynomial representations},  Academic Press Inc. (1971).
 

\bibitem{purcell} E. Purcell, {\it  Electricity and magnetism}, Mc Graw-Hill (1985).

\bibitem{schwartz} M. Schwartz, {\it Principles of electrodynamics}, Dover Publications (1987).

\bibitem{rmc} J. Reitz, F. Milford, R. Christy, {\it Foundations of Electromagnetic Theory}, Addison Wesley (1992).


\bibitem{reason} Such is the essence of reasoning that I find in some classic texts e.g. on page no. 45 of ref.\cite{maxwell} or on page no.24 of ref.\cite{jackson}.


\bibitem{jackson}J. D. Jackson, {\it Classical electrodynamics}, Wiley (1999).


\bibitem{millikan}R. Millikan, {\it On the elementary electrical charge and the avogadro constant}, Phys. Rev. 2, 109 (1913), \href{https://journals.aps.org/pr/abstract/10.1103/PhysRev.2.109}{online link}.

\bibitem{fletcher} H. Fletcher, {\it My Work with Millikan on the Oil-drop Experiment},  Physics Today 35, 6, 43 (1982), \href{https://physicstoday.scitation.org/doi/10.1063/1.2915126}{online link}.


\bibitem{rottenqed} O. Consa, {\it Something is rotten in the state of QED}, \href{https://vixra.org/abs/2002.0011}{vixra 2002.0011}.


\bibitem{eingr}A. Einstein, {\it The Formal Foundation of the General Theory of Relativity} in {\it  The Collected Papers of Albert Einstein,  Vol. 6,  The Berlin Years Writings, 1914-1917}, Princeton University Press (1997).




\bibitem{aryabhatiyaclark} W. Clark, {\it The Aryabhatiya of Aryabhata - an ancient Indian work on mathematics and astronomy}, The University of Chicago Press (1930).





\bibitem{yuktibhasa}K.V.Sarma, K.Ramasubramanian, M.D.Srinivas, M.S,Sriram, {\it Ganita-Yukti-Bhasa (Rationales in Mathematical Astronomy) of Jyeshthadeva, Volume 1- Mathematics, Volume 2- Astronomy}, Hindustan Book Agency (2008).

 
 \bibitem{calculusindia} P. P. Divakaran, {\it The first textbook of calculus: Yuktibhasa}, Indian Journal of Philosophy,  35:417-443 (2007).
 
 
 \bibitem{robinsonmeta} A. Robinson, {\it The Metaphysics of Calculus}, Studies in Logic and the Foundations of Mathematics, Vol. 47, pp. 28-46 (1967).
 
 \bibitem{robinson}A. Robinson, {\it Non-standard analysis}, North Holland Publishing Company, Amsterdam (1966).

\bibitem{companion} A. Majhi, {\it Logic, Philosophy and Physics: a critical commentary on the dilemma of categories}, Axiomathes (2021), \href{https://www.researchgate.net/publication/354332415_Logic_Philosophy_and_Physics_A_Critical_Commentary_on_the_Dilemma_of_Categories}{online link}.


\end{thebibliography}
\end{document}